\documentclass[11pt,letterpaper]{article}

\usepackage{graphicx}
\usepackage{subcaption}
\usepackage{booktabs}
\usepackage{tabularray}
\UseTblrLibrary{booktabs, siunitx}

\usepackage{amsmath}
\usepackage{amsthm}
\IfFileExists{keytheorems.sty}{\usepackage[thmtools-compat]{keytheorems}}{\usepackage{thmtools}}
\usepackage{amssymb}
\usepackage{mathrsfs}
\usepackage{mathtools}
\usepackage{bm}
\usepackage{nicefrac}
\usepackage{dsfont}

\usepackage{siunitx}
\usepackage[dvipsnames]{xcolor}

\usepackage{newfloat}
\DeclareFloatingEnvironment[fileext=lop, placement=htbp, name=LP]{linprog}
\DeclareFloatingEnvironment[fileext=iop, placement=htbp, name=IP]{intprog}

\usepackage{algorithm}
\usepackage{algpseudocode}

\usepackage{xurl}
\usepackage[backref=page,colorlinks=true]{hyperref}
\hypersetup{
    linkcolor=blue,
    citecolor=Plum,
    urlcolor=blue,
}
\usepackage[capitalise,nameinlink]{cleveref}

\usepackage[shortlabels]{enumitem}
\setlist[enumerate]{noitemsep,topsep=0pt,parsep=0pt,partopsep=0pt}
\setlist[itemize]{noitemsep,topsep=0pt,parsep=0pt,partopsep=0pt}

\usepackage{xspace}
\usepackage{fontawesome5}
\usepackage{soul}

\usepackage{tikz}
\usetikzlibrary{
    positioning,
    shapes,
    arrows.meta,
    calc,
    patterns.meta,
    patterns,
    decorations.pathreplacing,
    backgrounds,
    tikzmark,
    fit,
}

\colorlet{mmcclient}{blue!65!black}
\colorlet{mmcclientfill}{blue!8}
\colorlet{mmcfacility}{green!45!black}
\colorlet{mmcfacilityfill}{green!8}
\colorlet{mmcsteiner}{black!65}
\colorlet{mmcsteinerfill}{black!4}

\colorlet{mmcedge}{black!70}
\colorlet{mmcedgefaint}{black!45}
\colorlet{mmcgroup}{black!45}
\colorlet{mmctext}{black!70}

\definecolor{mmcaccentone}{RGB}{145, 70, 150}
\definecolor{mmcaccenttwo}{RGB}{214, 137, 16}

\newlength{\mmclinewidth}
\newlength{\mmcthicklinewidth}
\tikzset{
    mmc/client/.style={
            circle,
            draw=mmcclient,
            fill=mmcclientfill,
            line width=\mmclinewidth,
            minimum size=8mm,
            inner sep=1pt,
            font=\small
        },
    mmc/center/.style={
            rectangle,
            rounded corners=2pt,
            draw=mmcfacility,
            fill=mmcfacilityfill,
            line width=\mmclinewidth,
            minimum width=10mm,
            minimum height=7mm,
            inner sep=2pt,
            font=\small
        },
    mmc/steiner/.style={
            circle,
            draw=mmcsteiner,
            fill=mmcsteinerfill,
            line width=\mmclinewidth,
            minimum size=8mm,
            inner sep=1pt,
            font=\small
        },
    mmc/edge/.style={
            draw=mmcedge,
            line width=\mmclinewidth
        },
    mmc/edge light/.style={
            draw=mmcedgefaint,
            line width=\mmclinewidth
        },
    mmc/leader/.style={
            draw=mmcgroup,
            line width=0.6pt,
            densely dashed
        },
    mmc/arrow/.style={
            mmc/edge,
                -{Latex[length=2.4mm, width=1.8mm]}
        },
    mmc/accent one/.style={
            draw=mmcaccentone,
            line width=\mmcthicklinewidth,
            -{Latex[length=2mm, width=1.6mm]}
        },
    mmc/accent two/.style={
            draw=mmcaccenttwo,
            line width=\mmcthicklinewidth,
            -{Latex[length=2mm, width=1.6mm]}
        },
    mmc/edge label/.style={
            font=\scriptsize,
            fill=white,
            inner sep=2pt
        },
    mmc/note/.style={
            font=\scriptsize,
            text=mmctext
        },
    mmc/length/.style={
            font=\scriptsize
        },
    mmc/divider/.style={
            draw=mmcgroup!60,
            line width=0.8pt
        },
    mmc/group/.style={
            draw=mmcgroup,
            dashed,
            line width=0.8pt,
            rounded corners=4pt
        },
    mmc/stage/.style={
            draw=mmcclient,
            fill=mmcclientfill,
            rounded corners=2pt,
            line width=\mmclinewidth,
            align=left,
            text width=84mm,
            inner xsep=6pt,
            inner ysep=5pt,
            font=\small
        },
    mmc/stage ours/.style={
            mmc/stage,
            draw=mmcfacility,
            fill=mmcfacilityfill
        },
    mmc/stage arrow/.style={
            mmc/arrow,
            line width=\mmcthicklinewidth
        },
    mmc/stage note/.style={
            font=\scriptsize,
            text=mmctext,
            anchor=west,
            inner sep=2pt
        },
}

\DeclarePairedDelimiter{\abs}{\lvert}{\rvert}
\DeclarePairedDelimiter{\norm}{\lVert}{\rVert}

\DeclarePairedDelimiter{\set}{\{}{\}}

\DeclarePairedDelimiter{\bracket}{[}{]}

\DeclareMathOperator{\intr}{int}

\DeclareMathOperator{\supp}{supp}

\DeclareMathOperator*{\Expectation}{\mathbb{E}}
\newcommand*{\EE}[2][]{\Expectation_{#1}\bracket*{#2}}

\DeclareMathOperator*{\Probability}{\mathbb{P}}
\renewcommand*{\Pr}[2][]{\Probability_{#1}\bracket*{#2}}

\newcommand*{\one}[1]{\mathds{1}\bracket*{#1}}

\newcommand*{\dif}{\mathop{}\!\mathrm{d}}

\newcommand*{\RR}{\mathbb{R}}

\newcommand*{\calO}{\mathcal{O}}
\newcommand*{\calP}{\mathcal{P}}

\newcommand{\OPT}{\ensuremath{\mathsf{OPT}}\xspace}

\NewDocumentCommand{\cost}{o g}{%
    \ensuremath{\mathsf{cost}%
        \IfValueT{#1}{_{#1}}%
        \IfValueT{#2}{\!\left(#2\right)}%
    }\xspace%
}
\newcommand{\LP}{\ensuremath{\mathsf{LP}}\xspace}

\DeclareMathOperator{\divergence}{div}

\declaretheorem[name=Theorem, style=plain, numberwithin=section]{theorem}
\declaretheorem[name=Lemma, style=plain, sibling=theorem]{lemma}
\declaretheorem[name=Proposition, style=plain, sibling=theorem]{proposition}
\declaretheorem[name=Corollary, style=plain, sibling=theorem]{corollary}
\declaretheorem[name=Definition, style=definition, sibling=theorem]{definition}

\declaretheorem[name=Fact, style=plain, sibling=theorem, refname={Fact,Facts}, Refname={Fact,Facts}]{fact}

\DeclareMathOperator{\conv}{conv}

\newcommand{\cond}[1]{\textup{(\textsf{#1})}}
\newcommand{\gapvalue}{\frac{159389999561650025}{129321858575204877}}

\usepackage[margin=1in]{geometry}
\usepackage[T1]{fontenc}
\usepackage[final]{microtype}
\usepackage[numbers,square,sort&compress]{natbib}

\newcommand{\paragraphname}[2][0.75em]{\vspace{#1}\noindent\textbf{#2}}

\hypersetup{pdftitle={An Integrality-Gap Lower Bound of 1.2325 for Minimum Multiway Cut}}

\title{An Integrality-Gap Lower Bound of $1.2325$ for \\ Minimum Multiway Cut}

\usepackage{authblk}
\author{Dimitar Chakarov\thanks{\texttt{chakarov@ttic.edu}}}
\author{Yury Makarychev\thanks{\texttt{yury@ttic.edu}. Supported in part by NSF award ECCS-2216899.}}
\affil{Toyota Technological Institute at Chicago}

\date{}

\allowdisplaybreaks
\begin{document}
\maketitle

\begin{abstract}
    We prove that the integrality gap of the C\u{a}linescu--Karloff--Rabani relaxation for Minimum Multiway Cut is at least $1.2325$ when the number of terminals is large.
    By a result of Manokaran, Naor, Raghavendra, and Schwartz~(2008), it is therefore NP-hard, under the Unique Games Conjecture, to approximate Minimum Multiway Cut within a factor of $1.2325$.
    This improves the previously best known lower bound of $1.20016$ due to B\'erczi, Chandrasekaran, Kir\'aly, and Madan~(2020).
\end{abstract}

\section{Introduction}\label{sec:intro}

Multiway Cut is the problem of separating $k$ terminals in a weighted graph by removing edges of minimum total weight.
We are given an undirected graph $G = (V, E)$ with nonnegative edge weights $w_{uv}$ and a set of $k$ terminals $t_1, \ldots, t_k \in V$.
A \emph{$k$-way cut} is a labeling $C : V \to [k]$ with $C(t_i) = i$ for $i \in [k]$, and its \emph{cost} is the weight of the edges whose endpoints receive different labels,
\begin{equation}
    \cost{C} = \sum_{uv \in E} w_{uv} \, \one{C(u) \ne C(v)}.
\end{equation}
We write $\OPT$ for the minimum cost of a $k$-way cut.

When $k = 2$, Multiway Cut is the Minimum $s$--$t$ Cut problem, which is solvable in polynomial time.
However, Dahlhaus, Johnson, Papadimitriou, Seymour, and Yannakakis~\cite{DJPSY92,DJPSY94} showed that the problem is APX-hard for every fixed $k \ge 3$.
They also gave a simple combinatorial $(2 - 2/k)$-approximation algorithm for the problem.
A major advance was made by C\u{a}linescu, Karloff, and Rabani~\cite{CKR00}, who introduced a geometric approach to Multiway Cut based on linear programming.
Their relaxation, now commonly known as the CKR relaxation, is defined as follows:
\begin{equation}\label{eq:ckr}
    \begin{aligned}
        \text{minimize}\ \  & \sum_{uv \in E} w_{uv} \, d(x^u, x^v) \\
        \text{subject to}\ \  & x^u \in \Delta_k &  & \text{for every } u \in V, \\
         & x^{t_i} = e_i &  & \text{for every } i \in [k].
    \end{aligned}
\end{equation}
Here $\Delta_k = \set{x \in \RR^k_{\ge 0} : x_1 + \cdots + x_k = 1}$ is the simplex, $e_i$ is the $i$th vertex, and we use the $\ell_1$ metric scaled by $1/2$, i.e., $d(x, y) = \frac{1}{2} \norm{x - y}_1$.
Every $k$-way cut $C$ gives a feasible solution $x^u = e_{C(u)}$ of the same cost, so the optimal value $\LP$ of~\eqref{eq:ckr} is at most $\OPT$.
C\u{a}linescu, Karloff, and Rabani rounded their relaxation to obtain a $(3/2 - 1/k)$-approximation, and every approximation algorithm since rounds the same relaxation with a different rounding procedure~\cite{KKSTY04,BNS18,SV14,BHPZ26}.
In particular, Karger, Klein, Stein, Thorup, and Young~\cite{KKSTY04} settled the case of $k = 3$ with approximation ratio $12/11$, and $1.3438$ for general $k$.
Buchbinder, Naor, and Schwartz~\cite{BNS18} introduced the method of exponential clocks and designed a rounding scheme with ratio $1.3239$, and Sharma and Vondr\'ak~\cite{SV14} obtain a $1.2965$-approximation with a mixture of rounding schemes.
Buchbinder, Schwartz, and Weizman~\cite{buchbinder2017simplex} gave a simpler rounding scheme nearly matching the approximation guarantee of~\cite{SV14}.
The current best approximation ratio for large $k$ is $1.2787$ due to Brakensiek, Huang, Potechin, and Zwick~\cite{BHPZ26}.

Recall that the \emph{integrality gap} of an instance is the ratio $\OPT/\LP$.
Let $G_k$ denote the supremum of the integrality gaps over all instances with $k$ terminals.
Manokaran, Naor, Raghavendra, and Schwartz~\cite{MNRS08} showed that, assuming the Unique Games Conjecture, the best approximation ratio achievable in polynomial time for Minimum Multiway Cut with $k$ terminals equals $G_k$.

Compared to algorithmic results, lower bounds on the integrality gap are much less developed.
The gap is known exactly only for $k = 3$, where it is $12/11$~\cite{KKSTY04,CT99}.
For general $k$, the known lower bounds analyze a low-dimensional skeleton of the high-dimensional simplex~$\Delta_k$.
The $d$-dimensional skeleton is the union of the faces of dimension at most~$d$, that is, the points of~$\Delta_k$ with at most $d + 1$ positive coordinates.
Freund and Karloff~\cite{FK00} analyzed the one-dimensional skeleton and gave an asymptotic lower bound of $8/7$.
Angelidakis, Makarychev, and Manurangsi~\cite{AMM17} introduced the technique of non-opposite cuts, which makes it possible to analyze a low-dimensional skeleton of $\Delta_k$ without considering the entire simplex, and improved the bound to $1.2$ on the two-dimensional skeleton.
The current best lower bound of $1.20016$ is due to B\'erczi, Chandrasekaran, Kir\'aly, and Madan~\cite{BCKM20}.

\paragraphname{Our contribution.}
We obtain the following asymptotic lower bound on the integrality gap for large~$k$.

\begin{theorem}\label{thm:main}
    Let $G_k$ denote the supremum of the integrality gap of the CKR relaxation \eqref{eq:ckr} over all instances with $k$ terminals.
    Then, we have\footnote{We note that $\lim_{k \to \infty} G_k$ exists.
        The sequence $G_k$ is nondecreasing because adding an isolated terminal to an instance changes neither $\OPT$ nor $\LP$, and it is bounded by $3/2$, the approximation factor of the rounding of~\cite{CKR00}.
    }
    \[
        \lim_{k \to \infty} G_k \ge \gapvalue > 1.2325.
    \]
    Equivalently, for every $\eta > 0$ and all sufficiently large $k$, there is a finite $k$-terminal instance whose integrality gap is at least the constant above minus $\eta$.
\end{theorem}
This result combined with the hardness result of \cite{MNRS08}, immediately yields the following corollary.

\begin{corollary}\label{cor:ugc}
    Assuming the Unique Games Conjecture, for every $\varepsilon > 0$ and all sufficiently large $k$, it is NP-hard to approximate Multiway Cut with $k$ terminals within a factor $1.2325 - \varepsilon$.
\end{corollary}

The gap in \cref{thm:main} is in part obtained with a computer-assisted search via linear programming.
However, we can get the slightly weaker gap of $1.222376$ without resorting to numerical optimization tools~(\cref{thm:explicit}).
This is still better than the previous best lower bound of $1.20016$~\cite{BCKM20}.

\paragraphname{Our approach.}
A rounding scheme for~\eqref{eq:ckr} is a random $k$-way cut of the simplex itself, and by~\cite{KKSTY04} the gap $G_k$ is exactly the best approximation factor of such a cut, so a lower bound on the gap is a statement about every random $k$-way cut of $\Delta_k$.
To remove the dependence on $k$, we follow Angelidakis, Makarychev, and Manurangsi~\cite{AMM17} and restrict the cut to a face with a fixed number $q$ of vertices.
Up to an error that vanishes as $k$ grows, this turns it into a \emph{non-opposite cut} of $\Delta_q$.
Such a cut has one extra label $\ast$, which stands for all terminals outside the face, and every point receives a label from its support or $\ast$.
A lower bound on the approximation factor of every random non-opposite cut of $\Delta_q$ is then a lower bound on $\lim_{k \to \infty} G_k$.

On the faces of $\Delta_q$ we build a Multiway Cut instance together with label flows, one per label, fixed before the cut.
A cut selects at every point the flow of the label it assigns there, and the flows satisfy a conservation and a capacity condition, as in max-flow min-cut.
Conservation makes the flow of a label carry the weight of that label's terminal edges, so a cut that keeps a terminal edge must let that flow into the region of its label, across a boundary where the label changes.
Capacity makes the edges of the instance crossing that boundary weigh at least the difference of the two flows there.

Hence, a random non-opposite cut with finite approximation factor costs, in expectation, nearly as much as the cut that labels every point by $\ast$, and the ratio of this cost to the fractional cost of the instance is a lower bound on its approximation factor~(\cref{thm:general}).
We prove this for every $q$, so each explicit bound in \cref{thm:main,thm:explicit} is a single flow instance, on $\Delta_{11}$ and on $\Delta_{10}$, respectively.

\paragraphname{Roadmap.}
We introduce the notion of label flows and give an overview of our approach in \Cref{sec:overview}.
In \cref{sec:triangle} we exhibit a flow instance for $q = 3$.
\cref{sec:proof} proves the lower bound formally.
\cref{sec:explicit,sec:recipe} give the two flow instances on $\Delta_{10}$ and $\Delta_{11}$, respectively.

\section{Preliminaries on Non-Opposite Cuts}
\label{sec:reduction}

In this section we give an overview of the non-opposite cuts of \cite{AMM17}, and explain why we can restrict our attention to them.

A solution of~\eqref{eq:ckr} places each vertex $u$ at a point $x^u \in \Delta_k$, and a random $k$-way cut of the simplex, that is, a random labeling $C : \Delta_k \to [k]$ with $C(e_i) = i$ for all $i \in [k]$, rounds it by labeling $u$ with $C(x^u)$.
Such a cut has \emph{approximation factor} $\alpha$ if
\begin{equation}\label{eq:maxdensity}
    \Pr{C(x) \ne C(y)} \le \alpha \, d(x, y) \text{ for all } x, y \in \Delta_k.
\end{equation}
Then every edge $uv$ is cut with probability at most $\alpha \, d(x^u, x^v)$, so the rounding is an $\alpha$-approximation and $G_k \le \alpha$.
Karger, Klein, Stein, Thorup, and Young~\cite{KKSTY04} showed that some random $k$-way cut of $\Delta_k$ has approximation factor $G_k$.

To remove the dependence on $k$, Angelidakis, Makarychev, and Manurangsi~\cite[Lemma~13 of full version]{AMM17} restrict the cut to the face spanned by $q$ terminals chosen at random, which we identify with $\Delta_q$, and replace the labels of all other terminals by a single label $\ast$.
We do not treat $\ast$ as another terminal, so a point labeled $\ast$ has all of its terminal edges cut.
A point $x$ of the face rarely receives the label of a chosen terminal outside $\supp(x) = \set{i : x_i > 0}$, so replacing such labels by $\ast$ changes the approximation factor by an error that vanishes as $k \to \infty$.
The result is a cut of the following kind.

\begin{definition}[Non-opposite cuts]\label{def:nonopposite}
    A \emph{non-opposite cut} of $\Delta_q$ is a labeling $C : \Delta_q \to [q] \cup \set{\ast}$ with $C(e_i) = i$ for every $i \in [q]$ and $C(x) \in \supp(x) \cup \set{\ast}$ for every $x \in \Delta_q$.
    A \emph{random non-opposite cut} is a probability distribution over non-opposite cuts.
\end{definition}

For example, the side $x_1 = 0$ of $\Delta_3$ allows the labels $2$, $3$, and $\ast$ but forbids $1$, so no point of the side opposite $e_1$ receives its label.
The approximation factor of a random non-opposite cut is defined by~\eqref{eq:maxdensity} with $\Delta_q$ in place of $\Delta_k$.
In the limit $k \to \infty$ we obtain the following lemma, which we prove in \cref{app:reduction}.

\begin{lemma}[Non-opposite reduction]\label{lem:reduction}
    Fix $q \ge 2$ and $\beta > 0$.
    If every random non-opposite cut of~$\Delta_q$ with approximation factor $\alpha$ satisfies $\alpha \ge \beta$, then $\lim_{k \to \infty} G_k \ge \beta$, and for every $\eta > 0$ and all sufficiently large $k$ there is a finite $k$-terminal instance whose integrality gap is at least $\beta - \eta$.
\end{lemma}

\section{Technical Overview of the Approach}\label{sec:overview}

By \cref{lem:reduction}, it suffices to fix $q$ and show that every random non-opposite cut of $\Delta_q$ has approximation factor at least $\beta$.
The argument below works for every $q$, and \cref{thm:main} is the case $q = 11$.

Let $C$ be a random non-opposite cut of $\Delta_q$ with approximation factor $\alpha$, and consider a Multiway Cut instance whose vertices are points of $\Delta_q$ and whose terminals are the vertices $e_1, \ldots, e_q$.
Since $C$ severs an edge of length $d$ with probability at most~$\alpha d$, the expected cost of $C$ on the instance is at most $\alpha L$, where $L$ is the cost of the fractional solution that places every vertex at its own point.
Hence, if every such $C$ has expected cost at least~$M$ on the instance, then $\alpha \ge M/L$.
Our task is therefore to construct an instance together with a proof that every random non-opposite cut has expected cost at least $M$ on it, with $M/L$ as large as possible.
We obtain such proofs by designing flows, one for each label, that are fixed before the cut, and we call an instance together with these flows a \emph{flow instance}.

A flow instance determines three quantities: (1) $L$ is the fractional cost above, (2) $B$ is the cost of the cut that assigns the label $\ast$ to every point, and (3) $R$ is an allowance, a slack in the flow conditions.
Our main technical result, \cref{thm:general}, states that every random non-opposite cut of~$\Delta_q$ with approximation factor $\alpha$ satisfies $B - R \le \EE{\cost{C}} \le \alpha L$, and therefore $\alpha \ge (B - R)/L$.
So, up to the allowance, no non-opposite cut is cheaper in expectation than labeling every point by~$\ast$.
We first define flow instances on a finite graph (\cref{sec:overview-graph}), where the argument cannot give a bound above one, because it also applies to the fractional solution.
We then move to the faces of the simplex (\cref{sec:overview-simplex}), where it can.
Finally, we state \cref{thm:general} and reduce it to three lemmas (\cref{sec:overview-proof}), and describe the two flow instances behind \cref{thm:explicit,thm:main} (\cref{sec:overview-instances}).

\subsection{Flow Instances on a Graph}\label{sec:overview-graph}
Let $H$ be a finite graph whose vertices are points of $\Delta_q$, with edge weights $w_{uv} \ge 0$, in which every vertex $v$ also has a terminal edge of weight $g_{v, i} \ge 0$ to $e_i$ for each $i \in \supp(v)$.
Restricted to the vertices, a non-opposite cut $C$ of $\Delta_q$ gives every vertex $v$ a label in $A(v) = \supp(v) \cup \set{\ast}$.
These are the labels \emph{allowed} at $v$, while the labels $i \notin A(v)$, those with $v_i = 0$, are \emph{forbidden} at $v$.
The cost of $C$ is
\begin{equation}\label{eq:graphcost}
    \cost{C} = \sum_{uv} w_{uv} \, \one{C(u) \ne C(v)} + \sum_{v, i} g_{v, i} \one{C(v) \ne i}.
\end{equation}
Labeling every vertex of $H$ by $\ast$ severs exactly the terminal edges and costs $B = \sum_{v, i} g_{v, i}$, and the fractional cost of $H$ is $L = \sum_{uv} w_{uv} \, d(u, v) + \sum_{v, i} g_{v, i} (1 - v_i)$.
For every label $\ell$, a \emph{label flow} is a function $\varphi_\ell$ on the oriented edges of $H$ with $\varphi_\ell(v \to u) = - \varphi_\ell(u \to v)$, and $\varphi_\ast = 0$.
Let $\operatorname{in}_\ell(v)$ denote the net inflow of $\varphi_\ell$ at $v$.

We say that label flows together with allowances $\epsilon_v \ge 0$ form a \emph{flow instance} on $H$ if they satisfy the following two conditions.
\begin{description}[nosep]
    \item[\cond{Capacity}] On every edge $uv$, any two label flows differ by at most the weight of the edge, that is, $\abs{\varphi_\ell(u \to v) - \varphi_m(u \to v)} \le w_{uv}$ for all $\ell, m \in A(u) \cup A(v)$.
    \item[\cond{Conservation}] At every vertex $v$, the inflow of every allowed label $\ell \in A(v)$ covers the terminal edge of that label up to the allowance, that is, $\operatorname{in}_\ell(v) \ge g_{v, \ell} - \epsilon_v$.
\end{description}
We use $R = \sum_v \epsilon_v$ to denote the total allowance.
Nothing constrains the inflow of a forbidden label, so it may be negative, and such a vertex may emit flow of that label.
These vertices are where the label flows start, at no cost, because a cut never selects a forbidden label and conservation is never needed there.

\begin{lemma}[Flow lower bound on a graph]\label{lem:discrete}
    Every non-opposite cut of $H$ costs at least $B - R$.
    Consequently, if $L > 0$, every random non-opposite cut of $\Delta_q$ with approximation factor $\alpha$ satisfies $\alpha \ge (B - R)/L$.
\end{lemma}

Summing the inflow of the label $C(v)$ over all vertices $v$, conservation makes the sum at least the weight of the terminal edges that $C$ does not cut minus $R$, while an edge contributes the difference of two label flows if it is cut and nothing otherwise, so the sum is at most the weight of the cut edges.
The same summation applies to the fractional solution in which $v$ has label $i$ with weight $v_i$, and that solution costs $L$.

\begin{proposition}\label{prop:discretelimit}
    Every flow instance on a finite graph has $B - R \le L$, so $(B - R) / L \leq 1$.
    Hence, the bound $\alpha \ge (B - R)/L$ is trivial because we always have $\alpha \ge 1$.
\end{proposition}

We prove \cref{lem:discrete,prop:discretelimit} in \cref{sec:graphflows}.

\subsection{Flow Instances on the Simplex}\label{sec:overview-simplex}

We replace the finite graph by the truncated faces of $\Delta_q$, whose edges are infinitesimal.
The summation over vertices becomes a summation over faces, which still applies to the fractional solution, but capacity is now used only on the surfaces along which the label of a cut changes, and the fractional solution has no such surfaces.

Fix a threshold $T \in (1/2, 1)$.
For a subset $S \subseteq [q]$ with $\abs{S} \ge 2$, let $\Delta_S = \set{x \in \Delta_q : \supp(x) \subseteq S}$ be the face of $\Delta_q$ spanned by the vertices $e_i$ with $i \in S$, and let $\Omega_S = \set{x \in \Delta_S : \max_{i \in S} x_i \le T}$ be the \emph{truncated face} for $S$.
The boundary of $\Omega_S$ has two kinds of facets.
The \emph{terminal facet} $\Phi^S_i = \set{x \in \Omega_S : x_i = T}$, for $i \in S$, is where the coordinate $i$ reaches the threshold, and it is the piece of $\Omega_S$ closest to $e_i$.
When $\abs{S} \ge 3$, the \emph{coordinate facet} $\set{x \in \Omega_S : x_j = 0}$, for $j \in S$, is the truncated face $\Omega_{S \setminus \set{j}}$ of the next lower dimension, so faces of consecutive dimensions share it.

\begin{figure}[ht]
    \centering
    \begin{subfigure}[t]{0.45\textwidth}
        \centering
        \begin{tikzpicture}[scale=0.9]
            \path[use as bounding box] (-2.45,-0.7) rectangle (2.65,3.35);
            \coordinate (t1) at (0,3);
            \coordinate (t2) at (-2,0);
            \coordinate (t3) at (2,0);
            \coordinate (a) at (-0.666667,2);
            \coordinate (b) at (0.666667,2);
            \coordinate (c) at (1.333333,1);
            \coordinate (d) at (0.666667,0);
            \coordinate (e) at (-0.666667,0);
            \coordinate (f) at (-1.333333,1);
            \coordinate (o) at (0,1);
            \coordinate (y) at (0,2);
            \fill[mmcsteinerfill] (a)--(b)--(c)--(d)--(e)--(f)--cycle;
            \draw[mmc/edge light] (t1)--(t2)--(t3)--cycle;
            \foreach \t in {0.15,0.5,0.85}{
                \draw[mmc/edge light] ($(a)!\t!(b)$)--(t1);
                \draw[mmc/edge light] ($(e)!\t!(f)$)--(t2);
                \draw[mmc/edge light] ($(c)!\t!(d)$)--(t3);
            }
            \draw[draw=mmcclient, line width=\mmcthicklinewidth] (a)--(b) (c)--(d) (e)--(f);
            \draw[draw=mmcfacility, line width=\mmcthicklinewidth] (a)--(f) (b)--(c) (d)--(e);
            \fill (y) circle[radius=1.4pt];
            \node[mmc/note, below] at (y) {$y$};
            \node[mmc/note, fill=mmcsteinerfill, inner sep=1pt] at (0,0.55) {$\Omega_{\{1, 2, 3\}}$};
            \node[above] at (t1) {$e_1$};
            \node[below left] at (t2) {$e_2$};
            \node[below right] at (t3) {$e_3$};
            \node[mmc/note, below] at (0,-0.12) {$x_1 = 0$: $\Omega_{\{2, 3\}}$};
            \node[mmc/note, anchor=west] at (0.7,2.55) {$\set{y, e_1}$};
            \draw[mmc/leader] (0.65,2.53)--(0,2.53);
            \node[mmc/note, anchor=west] at (0.95,1.96) {$x_1 = T$};
            \draw[mmc/leader] (0.9,1.96)--(0.55,2);
        \end{tikzpicture}
        \caption{Faces and terminal edges.}
        \label{fig:mesh}
    \end{subfigure}\hfill
    \begin{subfigure}[t]{0.45\textwidth}
        \centering
        \begin{tikzpicture}[scale=0.9]
            \path[use as bounding box] (-2.45,-0.7) rectangle (2.65,3.35);
            \coordinate (t1) at (0,3);
            \coordinate (t2) at (-2,0);
            \coordinate (t3) at (2,0);
            \coordinate (a) at (-0.666667,2);
            \coordinate (b) at (0.666667,2);
            \coordinate (c) at (1.333333,1);
            \coordinate (d) at (0.666667,0);
            \coordinate (e) at (-0.666667,0);
            \coordinate (f) at (-1.333333,1);
            \coordinate (o) at (0,1);
            \fill[mmcsteinerfill] (a)--(b)--(c)--(d)--(e)--(f)--cycle;
            \fill[mmcaccentone!12] (-1.067,1.4)--(a)--(b)--(1.067,1.4) .. controls (0.3,1.75) and (-0.5,0.95) .. cycle;
            \draw[mmc/edge light] (t1)--(t2)--(t3)--cycle;
            \foreach \t in {0.15,0.5,0.85}{
                \draw[mmc/edge light] ($(a)!\t!(b)$)--(t1);
            }
            \draw[draw=mmcclient, line width=\mmcthicklinewidth] (a)--(b) (c)--(d) (e)--(f);
            \draw[draw=mmcfacility, line width=\mmcthicklinewidth] (a)--(f) (b)--(c) (d)--(e);
            \foreach \x/\z in {-0.36/-0.58,-0.12/-0.2,0.12/0.2,0.36/0.58}{
                \draw[mmc/accent two, line width=\mmclinewidth] (\x,0.03) .. controls (\x,0.9) and (\z,1.0) .. (\z,2);
            }
            \draw[draw=mmcaccentone, line width=\mmcthicklinewidth] (-1.067,1.4) .. controls (-0.5,0.95) and (0.3,1.75) .. (1.067,1.4);
            \node[mmc/note, fill=mmcaccentone!12, inner sep=1pt] at (0,1.6) {$C = 1$};
            \node[mmc/note, fill=mmcsteinerfill, inner sep=1pt] at (0,0.45) {$C \in \set{2, 3, \ast}$};
            \node[above] at (t1) {$e_1$};
            \node[below left] at (t2) {$e_2$};
            \node[below right] at (t3) {$e_3$};
            \node[mmc/note, below] at (0,-0.12) {$x_1 = 0$: label $1$ is forbidden};
            \node[mmc/note, anchor=west] at (0.75,2.55) {edges to $e_1$};
            \node[mmc/note, anchor=east, text=mmcaccentone] at (-1.15,1.52) {cut};
        \end{tikzpicture}
        \caption{Label $1$ and a possible cut.}
        \label{fig:labelone}
    \end{subfigure}
    \caption{The geometry and a cut for $q = 3$.
        Blue segments are terminal facets, green segments are the three intervals, and gray links lead to terminals.
        In (b), the terminal edges that start in a region labeled $1$ are not cut, and the flow of label $1$ that leaves through them must enter the region.
        The curved arrows are schematic.
        All label flows are fixed before the cut.
        Changing the labeling selects among them without redesigning them.
    }
    \label{fig:regions}
\end{figure}
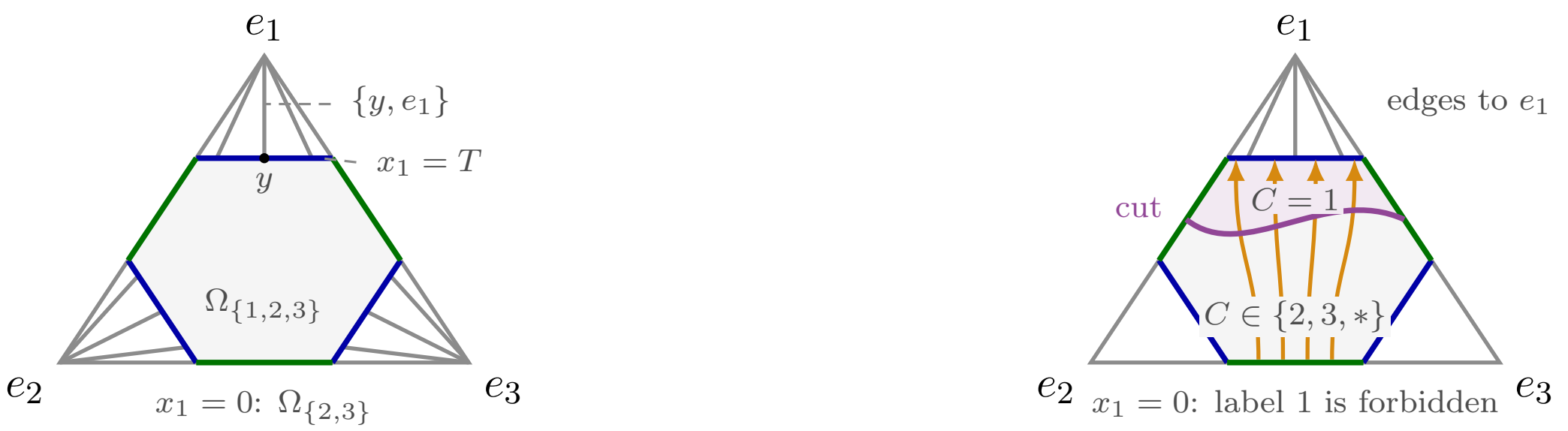

A \emph{weighting} of $\Delta_q$ consists of a weight density $w^S_{ij} : \Omega_S \to \RR_{\ge 0}$ for every $S$ and all $i < j$ in $S$, and a terminal density $g^S_i : \Phi^S_i \to \RR_{\ge 0}$ for every $S$ and $i \in S$.
The weighting plays the role of the weighted graph $H$, with densities in place of weights.
Inside $\Omega_S$, every point carries edges in each direction $e_i - e_j$, and $w^S_{ij}$ is their weight per unit cross-section.
Every point $y \in \Phi^S_i$ carries a terminal edge to $e_i$, of length $d(y, e_i) = 1 - T$, and $g^S_i$ is their weight per unit area.
\Cref{fig:mesh} shows the case $q = 3$, where $\Omega_{[3]}$ is a hexagon whose three coordinate facets are the intervals $\Omega_{\set{i, j}}$.
A weighting lives on all truncated faces at once, terminal edges start only on the terminal facets, and the corners $x_i > T$ are crossed only by terminal edges.

Restricted to $\Omega_S$, a non-opposite cut $C$ of $\Delta_q$ gives every point a label in $S \cup \set{\ast}$, the labels \emph{allowed} on $\Omega_S$, while the labels $j \notin S$ are \emph{forbidden} on $\Omega_S$.
In particular, label $j$ is forbidden on the coordinate facet $x_j = 0$.
One cut labels all of $\Delta_q$, so a point of the coordinate facet $x_j = 0$ has the same label in $\Omega_S$ and in $\Omega_{S \setminus \set{j}}$.
The cut severs an edge when its two endpoints have different labels, and its cost is
\begin{equation}\label{eq:cost}
    \cost{C} = \sum_{S} \sum_{i < j} \int_{\Omega_S} w^S_{ij} \dif \mu^S_{ij}(C) + \sum_{S} \sum_{i \in S} \int_{\Phi^S_i} g^S_i(y) \one{C(y) \ne i} \dif \sigma(y),
\end{equation}
where the \emph{crossing measure} $\mu^S_{ij}(C)(A)$ of a set $A \subseteq \Omega_S$ is the number of label changes of $C$ along a line in direction $e_i - e_j$ inside $A$, integrated over all such lines.
\Cref{lem:polyhedral} computes it for a cut whose label regions, the sets on which the cut takes one label, are polytopes, and \cref{sec:instance} makes it precise for measurable $C$.
Labeling every point of every truncated face by $\ast$ severs exactly the terminal edges and costs the total terminal weight
\begin{equation}\label{eq:B}
    B = \sum_{S} \sum_{i \in S} \int_{\Phi^S_i} g^S_i(y) \dif \sigma(y),
\end{equation}
and the solution of~\eqref{eq:ckr} that places every point at itself has the fractional cost
\begin{equation}\label{eq:L}
    L = \sum_{S} \sum_{i < j} \int_{\Omega_S} w^S_{ij}(x) \dif x + (1 - T) B.
\end{equation}

For every $S$ and every label $\ell \in S \cup \set{\ast}$, a \emph{label flow} is a vector field $F^S_\ell$ on $\Omega_S$, tangent to the face, with $F^S_\ast = 0$.
It plays the role of $\varphi_\ell$, with the flux of $F^S_\ell$ through a surface in place of $\varphi_\ell(u \to v)$, and $- \divergence F^S_\ell$, the net inflow per unit volume, in place of $\operatorname{in}_\ell(v)$.
It is piecewise smooth, i.e., $\Omega_S$ is cut into finitely many polytopes, the \emph{cells}, fixed before the cut, and $F^S_\ell$ is smooth on each closed cell.
Across a wall between two cells the two formulas may differ, but their flow must pass through the wall and none should be created on it.
We call this \emph{normal continuity}, and make it precise in \cref{sec:construction}.
The flow leaving $\Omega_S$ through the coordinate facet $x_j = 0$ is $- F^S_{\ell, j}$ per unit area, and through the terminal facet $\Phi^S_i$ it is $F^S_{\ell, i}$.
Each face has its own label flows, so the flow on $\Omega_S$ is not the restriction of a flow on a larger face, and the flows of different faces interact only through conservation.

\paragraphname{What a label flow does.}
The flow of label $j$ on $\Omega_S$ enters for free through the coordinate facet $x_j = 0$, where label $j$ is forbidden, and is created inside the face only from the flux arriving from larger faces or from the allowance.
It leaves through the terminal facet $\Phi^S_j$ at the terminal density $g^S_j$, carrying the weight of the terminal edges there, or through another coordinate facet $x_i = 0$ into a smaller face, where it continues as that face's flow of label $j$.
A cut selects at every point the flow of its label.
Where $C(y) = j$ on $\Phi^S_j$, the terminal edges are not cut, and the flow leaving through them must enter the region labeled $j$.
A non-opposite cut never labels a point of $x_j = 0$ by $j$, so the free entry is unavailable to that region, and the flow must cross the boundary of the region, where capacity charges it to the cost, or be paid for by conservation, through the larger faces and the allowance (\cref{fig:labelone}).
The three conditions below make this precise.

\begin{definition}[Flow instance on the simplex]\label{def:certificate}
    A \emph{flow instance} on $\Delta_q$ is a weighting together with label flows $F^S_\ell$ for every $S \subseteq [q]$ with $\abs{S} \ge 2$ and every $\ell \in S \cup \set{\ast}$, and allowances $\epsilon_S : \Omega_S \to \RR_{\ge 0}$ for every $S$, that satisfy the following three conditions for every $S$.
    \begin{description}[nosep]
        \item[\cond{Capacity}] Summed over any set of coordinates $I$, two label flows differ by at most the weight of the directions $e_i - e_j$ with $i \in I$ and $j \notin I$, that is, at every point $x$ of every cell of $\Omega_S$, all labels $\ell, m \in S \cup \set{\ast}$, and every nonempty proper $I \subset S$,
              \begin{equation}\label{eq:cap}
                  \abs*{\sum_{i \in I} \bigl( F^S_{\ell, i}(x) - F^S_{m, i}(x) \bigr)} \le \sum_{i \in I, \, j \in S \setminus I} w^S_{ij}(x).
              \end{equation}
        \item[\cond{Terminal condition}] Through the terminal facet $\Phi^S_i$, the flow of label $i$ leaves at the terminal density and no other label leaves, that is, $F^S_{\ell, i} = g^S_i \one{\ell = i}$ on $\Phi^S_i$ for every $i \in S$ and every label $\ell$.
        \item[\cond{Conservation}] On $\Omega_S$, the divergence of the flow of every allowed label is covered by the flux of the same label arriving through the coordinate facets of the faces one dimension up, up to the allowance, that is, for every $\ell \in S \cup \set{\ast}$,
              \begin{equation}\label{eq:cons}
                  \divergence F^S_\ell \le - \sum_{j \notin S} F^{S \cup \set{j}}_{\ell, j} + \epsilon_S.
              \end{equation}
    \end{description}
\end{definition}

We use $R = \sum_S \int_{\Omega_S} \epsilon_S \dif x$ to denote the total allowance and call $(B - R)/L$ the \emph{value} of the flow instance.

\subsection{The Main Technical Result}\label{sec:overview-proof}

\begin{theorem}[Flow lower bound]\label{thm:general}
    Let a flow instance on $\Delta_q$ have $L > 0$.
    Then every random non-opposite cut $C$ of $\Delta_q$ with approximation factor $\alpha$ satisfies
    \[
        B - R \le \EE{\cost{C}} \le \alpha L.
    \]
    Consequently, $\alpha \ge (B - R)/L$ and $\lim_{k \to \infty} G_k \ge (B - R)/L$.
\end{theorem}

We rewrite the cost~\eqref{eq:cost} as $\cost{C} = \sum_S \cost[S]{C} + \cost[\mathrm{term}]{C}$, where $\cost[S]{C}$ is the weight of the edges severed inside $\Omega_S$ and $\cost[\mathrm{term}]{C}$ is the weight of the cut terminal edges.
At a point $x$ of $\Omega_S$ the \emph{selected flow} is $F^S_{C(x)}(x)$, the flow of the label that $C$ assigns there.
The \emph{flux jump} $\Lambda_S(C)$ is the selected flow leaving $\Omega_S$ through its boundary minus the selected divergence inside,
\begin{equation}\label{eq:Lambda}
    \Lambda_S(C) = \sum_{i \in S} \int_{\Phi^S_i} F^S_{C(y), i}(y) \dif \sigma(y) - \sum_{j \in S} \int_{\Omega_{S \setminus \set{j}}} F^S_{C(x), j}(x) \dif x - \int_{\Omega_S} \divergence F^S_{C(x)}(x) \dif x,
\end{equation}
where $\divergence F^S_{C(x)}(x)$ is the divergence at $x$ of the label flow $F^S_\ell$ with $\ell = C(x)$, the two boundary integrals use the labels that $C$ assigns on the boundary, and the middle sum is absent when $\abs{S} = 2$.
If $C$ is constant on $\Omega_S$, the divergence theorem gives $\Lambda_S(C) = 0$.
In general the selected flow jumps where the label changes, and $\Lambda_S(C)$ tracks these jumps.

\begin{proof}[Proof of \cref{thm:general}, assuming \cref{lem:summation,lem:oneface,lem:expectedcost}]
    Let $C$ be a random non-opposite cut of $\Delta_q$ with approximation factor $\alpha$.
    All integrands in $\Lambda_S(C)$ and $\cost[\mathrm{term}]{C}$ are bounded, so their expectations exist, and
    \begin{align*}
        B - R
         & \le \sum_S \EE{\Lambda_S(C)} + \EE{\cost[\mathrm{term}]{C}}
         &  & \because\ \text{\cref{lem:summation}, for every realization of $C$} \\
         & \le \sum_S \EE{\cost[S]{C}} + \EE{\cost[\mathrm{term}]{C}} = \EE{\cost{C}}
         &  & \because\ \text{\cref{lem:oneface}, on every face} \\
         & \le \alpha L
         &  & \because\ \text{\cref{lem:expectedcost}}.
    \end{align*}
    Dividing by $L > 0$ gives $\alpha \ge (B - R)/L$.
    If $B - R \le 0$, then $\lim_{k \to \infty} G_k \ge (B - R)/L$ because $G_k \ge 1$, and otherwise \cref{lem:reduction} with $\beta = (B - R)/L$ gives it.
\end{proof}

We now explain intuitively why the three inequalities hold, and defer formal proofs to \cref{sec:proof}.
The first inequality holds for every cut and uses only the terminal condition and conservation.
By the terminal condition, the selected flow leaving through the terminal facets has the weight of the terminal edges that $C$ does not cut.
By conservation, the flow a face creates is at most what arrives from the larger faces, up to its allowance.
Summing over all faces, the flux jumps are at least the weight of the uncut terminal edges minus $R$, and adding the cut terminal edges gives $B - R$.

The second inequality uses the geometry of the cut, and unlike the first, it holds only in expectation.
We call the surfaces where the label of $C$ changes its \emph{interfaces}.
At an interface the selected flow jumps from one label flow to another, and capacity says that the cost of the interface is at least this jump, so the flux jump of a face is at most its cost.
When the label regions are polytopes, this follows from the divergence theorem applied to each region (\cref{lem:polyhedral}).
A measurable cut need not have interfaces in any usable sense, and this is where the approximation factor enters.
It implies that the label regions have finite perimeter almost surely (\cref{lem:variation}), so that their interfaces form a surface of finite area along which exactly two labels meet almost everywhere (\cref{fact:structure}), and it recovers in expectation the labels on the boundary of a face from the labels nearby (\cref{lem:boundaryflux}).

The third inequality is the approximation factor applied edge by edge to the weighting.

\subsection{Two Flow Instances}\label{sec:overview-instances}
The simplest flow instances have constant weights and affine label flows on every face, i.e., on every face the flow of label $i$ points toward $e_i$ from a point that depends affinely on $x$, with a scale that depends only on the dimension of the face.
For this family of instances we can compute $B$, $R$, and $L$ in terms of $q$ and $T$.
For the case $q = 10$ and $T = 3/5$ we obtain the following.

\begin{theorem}[Explicit flow instance]\label{thm:explicit}
    There is a flow instance on $\Delta_{10}$ with threshold $T = 3/5$, affine label flows and constant weights whose value is
    \[
        \frac{B - R}{L} = \frac{9999320418500}{8180229294583} = 1.222376\ldots.
    \]
\end{theorem}

See \cref{sec:explicit} for further details.

The conditions of \cref{def:certificate} are linear in the weights, terminal densities, label flows, and allowances, and the value is invariant under scaling them, so with $L = 1$ the flow instance of largest value in a finite-dimensional family is found by a linear program.
Our family is symmetric under permutations of the coordinates, subdivides the region of each face on which its coordinates are sorted into simplices, the \emph{cells}, and takes quadratic label flows and weights on every cell, in a basis whose nonnegativity turns inequalities between coefficients into inequalities at every point.
The resulting flow instance is given by a table of integers.
Normal continuity and the terminal condition follow from two boundary rules on the table, capacity from finitely many integer inequalities, and conservation from the definition of the allowance, so none of them depends on how the table was found (\cref{prop:recipe}).

\begin{proposition}[Computer-aided flow instance]\label{prop:certificate}
    There is a flow instance on $\Delta_{11}$ with $T = 249/400$ and value
    \[
        \frac{B - R}{L} = \gapvalue = 1.2325\ldots.
    \]
\end{proposition}

\Cref{thm:main} follows from \cref{lem:reduction,thm:general,prop:certificate}.
See \cref{sec:recipe,sec:recorded} for further details.

\section{An Example on the Triangle}\label{sec:triangle}

\begin{figure}[ht]
    \centering
    \begin{tikzpicture}[scale=0.9]
        \path[use as bounding box] (-2.45,-0.7) rectangle (2.65,3.35);
        \coordinate (t1) at (0,3);
        \coordinate (t2) at (-2,0);
        \coordinate (t3) at (2,0);
        \coordinate (a) at (-0.666667,2);
        \coordinate (b) at (0.666667,2);
        \coordinate (c) at (1.333333,1);
        \coordinate (d) at (0.666667,0);
        \coordinate (e) at (-0.666667,0);
        \coordinate (f) at (-1.333333,1);
        \fill[mmcsteinerfill] (a)--(b)--(c)--(d)--(e)--(f)--cycle;
        \draw[mmc/edge light] (t1)--(t2)--(t3)--cycle;
        \draw[draw=mmcclient, line width=\mmcthicklinewidth] (a)--(b) (c)--(d) (e)--(f);
        \draw[draw=mmcfacility, line width=\mmcthicklinewidth] (a)--(f) (b)--(c) (d)--(e);
        \draw[mmc/accent two] (0.95,1.6) .. controls (0,1.5) .. (-0.95,0.95);
        \draw[mmc/accent two] (1.05,1.35) .. controls (0.3,0.9) .. (-0.75,0.55);
        \draw[mmc/accent two] (1.1,1.15) .. controls (0.5,0.6) .. (0.2,0.06);
        \draw[mmc/arrow] (0.45,0.35)--(0.45,-0.12);
        \node[mmc/note, anchor=west, align=left, fill=white, inner sep=1pt] at (0.8,0.4) {outflow\\$-F_{2,1} = 2/3$};
        \draw[mmc/accent two] (0.4,-0.12)--(-0.45,-0.12);
        \draw[mmc/accent two] (-1.0,0.75)--(-1.6,0.45);
        \draw[mmc/accent two] (-0.55,-0.12)--(-1.2,-0.12);
        \node[above] at (t1) {$e_1$};
        \node[below left] at (t2) {$e_2$};
        \node[below right] at (t3) {$e_3$};
        \node[mmc/note, anchor=west, align=left] at (0.65,2.45) {$x_2 = 0$:\\ label $2$ forbidden};
        \node[mmc/note, below] at (0,-0.28) {$\Omega_{\{2, 3\}}$: $\divergence F_2 = 2/3$};
    \end{tikzpicture}
    \caption{Label $2$ and its flow into the interval $\Omega_{\set{2, 3}}$.
        The flow of label $2$ leaves through $x_1 = 0$ at density $-F_{2,1}^{[3]} = 2/3$, which is the divergence of the interval flow.
        The black arrow marks the flow leaving through $x_1 = 0$, and the curved arrows are schematic.
        The formulas in the text specify the label flows, and the flux values use the coordinate normalization, not distances in the drawing.
    }
    \label{fig:flowpicture}
\end{figure}
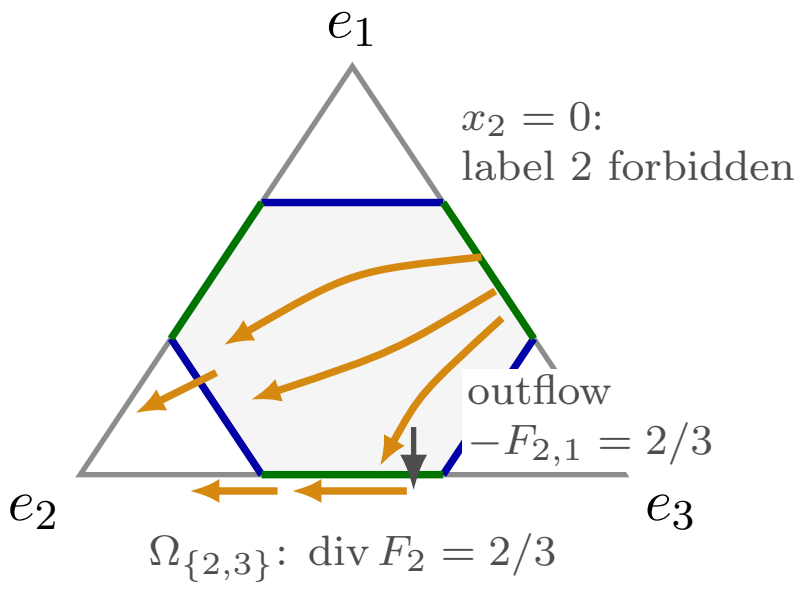

For $q = 3$ and $T = 2/3$ we give a flow instance on $\Delta_3$, an example of the objects defined in \cref{sec:overview}.
Then we explain why and how the three flow conditions hold, and compute the value of the flow instance.
The geometry is that of \cref{fig:regions}, and \cref{fig:flowpicture} shows the flow of label $2$.

\paragraphname{The geometry.}
We measure the hexagon in the coordinates $(x_1, x_2)$, with $x_3 = 1 - x_1 - x_2$, and each interval $\Omega_{\set{i, j}}$ in the coordinate $x_i$.
In these coordinates the triangle has vertices $(0, 0)$, $(1, 0)$, and $(0, 1)$ and area $1/2$.
Each removed corner has base and height $1/3$ and area $1/18$, so the hexagon has area $1/2 - 3 \cdot (1/18) = 1/3$.
On the interval $\Omega_{\set{i, j}}$ the coordinate $x_i$ runs from $1/3$ to $2/3$, so the interval has length $1/3$.
On the terminal facet $\Phi^{[3]}_i$, fixing $x_i = 2/3$ leaves $x_j + x_s = 1/3$, so the facet has length $1/3$ in the coordinate $x_j$.

\paragraphname{The flow instance.}
On the hexagon we take every weight to be $w^{[3]}_{st} = 1/2$, every terminal density to be $g^{[3]}_i = 1$, and the allowance to be $\epsilon_{[3]} = 2$.
The flow of a terminal label $i$ on the hexagon is
\[
    F^{[3]}_i(x) = e_i - p^{[3]}(x) \text{ with } p^{[3]}_s(x) = \frac{2}{3} - x_s \text{ for } s \in [3],
\]
and $F^{[3]}_\ast = 0$.
The coordinates of $p^{[3]}(x)$ are nonnegative and sum to one, so $p^{[3]}(x)$ is a point of the triangle, and the flow of label $i$ at $x$ points from this point toward the vertex $e_i$.
On each interval $\Omega_{\set{i, j}}$ we take the weight $w^{\set{i, j}}_{ij} = 2/9$, the terminal weights $g^{\set{i, j}}_i = g^{\set{i, j}}_j = 2/9$ at the two endpoints, the allowance $\epsilon_{\set{i, j}} = 0$, and the flows
\[
    F^{\set{i, j}}_i(x) = \frac{2}{3} \Bigl( x_i - \frac{1}{3} \Bigr) (e_i - e_j) \text{ and } F^{\set{i, j}}_j(x) = \frac{2}{3} \Bigl( x_j - \frac{1}{3} \Bigr) (e_j - e_i),
\]
with $F^{\set{i, j}}_\ast = 0$.
An endpoint carries counting measure, so its terminal density is a weight.
Each interval flow points along the interval toward its terminal and grows at rate $2/3$, its divergence.

\begin{proposition}[A flow instance on the triangle]\label{prop:triangle}
    The given flow instance on $\Delta_3$ has $B = 7/3$, $R = 2/3$, and $L = 3/2$.
    Consequently, every random non-opposite cut of $\Delta_3$ with approximation factor $\alpha$ satisfies $\alpha \ge 10/9$.
\end{proposition}

\Cref{fig:flowpicture} shows what the flow of label $2$ does on the hexagon.
On the side $x_2 = 0$, where label $2$ is forbidden, the second coordinate of $F^{[3]}_2$ is $1 - 2/3 = 1/3$, so the flow enters the hexagon there at density $1/3$.
On the terminal facet $\Phi^{[3]}_2$ we have $p^{[3]}_2 = 0$, so the second coordinate of $F^{[3]}_2$ is $1$ and the flow leaves at the terminal density $g^{[3]}_2 = 1$.
The second coordinates of $F^{[3]}_1$, $F^{[3]}_3$, and $F^{[3]}_\ast$ vanish there, which is the terminal condition.

On the side $x_1 = 0$ the first coordinate of $F^{[3]}_2$ is $- p^{[3]}_1 = - 2/3$, so the flow leaves the hexagon there at density $2/3$ and enters the interval $\Omega_{\set{2, 3}}$, where the flow $F^{\set{2, 3}}_2$ has divergence $2/3$.
This is conservation on the interval with zero allowance, and the same holds for every terminal label on every interval.
On the hexagon every terminal flow has divergence $2$, there is no face above the hexagon, and the allowance $\epsilon_{[3]} = 2$ covers it.
By symmetry the flow of label $2$ also leaves through $x_3 = 0$ into $\Omega_{\set{1, 2}}$ at density $2/3$.
Over the whole hexagon, entering plus created equals leaving,
\[
    \frac{1}{3} \cdot \frac{1}{3} + 2 \cdot \frac{1}{3} = \frac{7}{9} = 1 \cdot \frac{1}{3} + 2 \cdot \frac{2}{3} \cdot \frac{1}{3},
\]
which is $\Lambda_{[3]}(C) = 0$ for the cut $C \equiv 2$ of \cref{sec:overview-proof}.
\Cref{fig:labelone} shows a region labeled $1$ bounded by a cut.
On the cut the selected flow jumps from $F^{[3]}_1$ to the flow of the label on the other side, and capacity charges this jump to the cost of the cut.

Capacity on the hexagon is a check of the sums over the singletons $I = \set{i}$, since the complement of a singleton gives the same inequality for tangent flows.
The $i$th coordinate of $F^{[3]}_\ell$ is $1 - p^{[3]}_i$ for $\ell = i$, it is $- p^{[3]}_i$ for the other two terminal labels, and it is $0$ for $\ast$.
Since $0 \le p^{[3]}_i \le 2/3$ on the hexagon, any two of these values differ by at most $1$, and the weight of the two directions leaving $\set{i}$ is $1/2 + 1/2 = 1$.
On an interval the two terminal flows have $i$th coordinates $(2/3)(x_i - 1/3)$ and $(2/3)(x_i - 2/3)$, which differ by $2/9$, the weight of the interval, and the flow of $\ast$ lies between them.

We verify the flow conditions in \cref{app:triangle}.

\paragraphname{The value of the flow instance.}
The hexagon has three terminal facets of length $1/3$ with terminal density $1$, which contribute $3 \cdot (1/3) \cdot 1 = 1$ to $B$.
The three intervals have six endpoints in all, each with terminal weight $2/9$, which contribute $6 \cdot (2/9) = 4/3$, so $B = 7/3$.
The total allowance is the allowance $2$ on the hexagon times its area $1/3$, and the intervals contribute nothing, so $R = 2/3$.
There is no factor for the number of labels in $R$, because at every point the cut selects one flow, whose divergence is at most $2$.
The fractional cost adds, for every face and every pair $i < j$ of its coordinates, the weight $w^S_{ij}$ times the volume of the face, and then the terminal edge length $1/3$ times $B$.
The hexagon has three pairs and each interval has one, so
\[
    L = \underbrace{3 \cdot \frac{1}{2} \cdot \frac{1}{3}}_{\text{hexagon}} + \underbrace{3 \cdot \frac{2}{9} \cdot \frac{1}{3}}_{\text{intervals}} + \underbrace{\frac{1}{3} \cdot \frac{7}{3}}_{\text{terminal edges}} = \frac{3}{2}.
\]
The value of the flow instance is $(B - R)/L = (5/3)/(3/2) = 10/9$.

\paragraphname{Why the hexagon flows are not graph flows.}
On the hexagon any two terminal flows differ by $F^{[3]}_i - F^{[3]}_j = e_i - e_j$, and every weight is $1/2$.
On a graph whose edges at a point run in the three directions $e_s - e_t$, each with weight $1/2$, the difference of two label flows is a combination $a (e_1 - e_2) + b (e_1 - e_3) + c (e_2 - e_3)$ of the edge directions with $\abs{a}, \abs{b}, \abs{c} \le 1/2$.
Since $e_1 - e_2 = (e_1 - e_3) - (e_2 - e_3)$, the only such combination equal to $e_1 - e_2$ has $a = b = 1/2$ and $c = - 1/2$, and the only one equal to $e_1 - e_3$ has $a = b = c = 1/2$.
On a graph each label has one flow on each edge, so the difference for labels $2$ and $3$ is the difference of these two combinations, which sends $1$ along $e_2 - e_3$, twice the weight.
On the hexagon no such consistency is required.
Capacity constrains each pair of label flows on its own, in the normal direction of an interface, and an interface separates two labels at almost every point, so every pair may use the whole weight of every direction.
This is what \cref{prop:discretelimit} ruled out on the graph.

\section{Proof of the Flow Lower Bound}\label{sec:proof}

\Cref{sec:overview-proof} derived \cref{thm:general} from three lemmas, one for each inequality of its chain, and this section states and proves them.
\Cref{sec:construction} fixes what \cref{sec:overview} left informal, namely the measures and normals on the faces, the regularity of the label flows, and the form of capacity that holds for every normal.
\Cref{sec:summation} proves \cref{lem:summation}, the first inequality, and \cref{lem:polyhedral}, the second inequality for a cut whose label regions are polytopes.
\Cref{sec:instance} makes the crossing measure of \cref{sec:overview-simplex} precise and derives from the approximation factor the regularity that a measurable cut lacks, and \cref{sec:boundary} recovers the labels on the boundary of a face.
\Cref{sec:completing} then proves \cref{lem:oneface} and \cref{lem:expectedcost}.

Throughout, a random non-opposite cut is a map $C \colon \Xi \times \Delta_q \to [q] \cup \set{\ast}$ on a probability space $\Xi$ whose every realization $C(\xi, \cdot)$ is a non-opposite cut.
It is measurable for the product of the $\sigma$-algebra of $\Xi$ and the Borel sets of $\Delta_q$.
This is the precise meaning of \cref{def:nonopposite}, and the cuts of \cref{lem:reduction} are of this kind (\cref{app:reduction}).
Probabilities and expectations are over $\Xi$, and the geometry, the label flows, the weights, and every test function are fixed.

\subsection{Conventions and Precise Definitions}\label{sec:construction}

Fix $q \ge 2$ and the threshold $T \in (1/2, 1)$.
Since $1/\abs{S} \le 1/2 < T$, the truncated face $\Omega_S$ contains the barycenter of $\Delta_S$ and has dimension $\abs{S} - 1$, and since $T < 1$, it contains no vertex of the simplex.
Since $T > 1/2$, two terminal facets of the same face do not meet.
When $\abs{S} \ge 3$, the coordinate facet $x_j = 0$ of $\Omega_S$ is the truncated face $\Omega_{S \setminus \set{j}}$, because the truncation $\max_{i \in S} x_i \le T$ and the restriction $x_j = 0$ commute.
An interval $\Omega_{\set{i, j}}$ has two terminal facets, its endpoints $x_i = T$ and $x_j = T$, and no coordinate facet.

For $U = S \setminus \set{j}$ with $\abs{U} \ge 2$, we call $S = U \cup \set{j}$ a \emph{parent} of $U$ and $U$ a \emph{child} of $S$.
A face $U$ with $r$ coordinates has $q - r$ parents, one for each $j \notin U$, and $r$ children when $r \ge 3$.
The relation is not a rooted tree, since for $q = 4$ the interval $\Omega_{\set{1, 2}}$ has the two parents $\set{1, 2, 3}$ and $\set{1, 2, 4}$.
What a face shares with its children is only the cut, since one cut labels all of $\Delta_q$ and a point of a coordinate facet has one label.
The parent's flow through the shared facet is the supply of~\eqref{eq:supply} below.

The tangent space of $\Delta_S$ is $V_S = \set{v \in \RR^S : \sum_{i \in S} v_i = 0}$.
Delete any one coordinate to obtain independent coordinates on the affine hull of $\Delta_S$, and give the hull the Lebesgue volume measure $\dif x$ in these coordinates.
Changing the deleted coordinate is a linear map of determinant $\pm 1$, so $\dif x$ is well defined.
In the metric $d$ the coordinate directions have unit speed, $d(x, x + h(e_i - e_j)) = \abs{h}$.
For a full-dimensional polytope $A$ in the hull, take its outward unit normal and its surface measure $\dif \sigma$ in one such choice of coordinates, and write $\nu_A(F)$ for the normal component of a tangent vector field $F$.
It is a signed outward flux density, and its integral over a piece of $\partial A$ is the flux through that piece, positive when leaving $A$.

The divergence $\divergence F$ is the sum of the partial derivatives of the components of $F$ in the independent coordinates.
Since the components sum to zero, it does not depend on the deleted coordinate, and for $F$ continuously differentiable on $A$ the divergence theorem reads
\begin{equation}\label{eq:divthm}
    \int_A \divergence F \dif x = \int_{\partial A} \nu_A(F) \dif \sigma.
\end{equation}
The normal and the surface measure must use matching coordinates.
Although they separately depend on which coordinate is deleted, their product in this formula does not, and we always change the two together when rewriting a boundary integral.
Represent a normal covector by $\nu \in \RR^S$, modulo adding a constant to every coordinate, so that $\nu(e_i - e_j) = \nu_i - \nu_j$.

On a coordinate facet $x_j = 0$ of $\Omega_S$ keep $x_j$ among the independent coordinates, and on a terminal facet $x_i = T$ keep $x_i$.
The outward normals decrease $x_j$ and increase $x_i$, respectively, giving
\begin{equation}\label{eq:facetnormals}
    \nu_{\Omega_S}(F) = - F_j \text{ on } x_j = 0
    \text{ and }
    \nu_{\Omega_S}(F) = F_i \text{ on } x_i = T.
\end{equation}
With this choice the surface measure of the coordinate facet $x_j = 0$ is exactly the volume measure of the child $\Omega_{S \setminus \set{j}}$, so a parent's flux and a child's divergence combine without a scale factor.
On a terminal facet, fix $x_i = T$ and delete one more coordinate.
In every integral, $\dif \sigma$ denotes the surface measure on the indicated boundary or interface in matching coordinates, which is length on a two-dimensional face, area on a three-dimensional face, and counting measure at the endpoints of an interval.
We write $\sigma(J)$ for the measure of a set $J$.
By~\eqref{eq:facetnormals} the flux jump~\eqref{eq:Lambda} is
\[
    \Lambda_S(C) = \int_{\partial \Omega_S} \nu_{\Omega_S}\bigl(F^S_{C(y)}(y)\bigr) \dif \sigma(y) - \int_{\Omega_S} \divergence F^S_{C(x)}(x) \dif x,
\]
the selected outward flux through the whole boundary of $\Omega_S$ minus the selected divergence inside.

A label flow $F^S_\ell$ of \cref{sec:overview-simplex} is a bounded tangent vector field $F^S_\ell \colon \Omega_S \to V_S$ with the following regularity.
The face $\Omega_S$ carries a finite subdivision $\calP_S$ into polytopes of dimension $\abs{S} - 1$ with disjoint interiors, the \emph{cells}, fixed before the cut, and $Z_S$ denotes the union of the cell boundaries inside the interior of $\Omega_S$, a set of volume zero.
The flow is continuously differentiable on every closed cell, a point of a cell boundary being evaluated with the formula of the cell in question, and we write $F^S_{\ell, i}$ for its $i$th component.
\emph{Normal continuity} means that on every facet shared by two cells $P$ and $P'$, the normal component $\nu_P(F^S_\ell)$ computed from the formula on $P$ equals $- \nu_{P'}(F^S_\ell)$ computed from the formula on $P'$, for every label $\ell$.
The label regions of a cut form a different partition, which may cross cells freely.
We require of a weighting that each weight density $w^S_{ij} = w^S_{ji}$ is a bounded Borel function on $\Omega_S$ whose restriction to the interior of each cell extends continuously to the closed cell.
Each terminal density $g^S_i$ is a bounded measurable function on $\Phi^S_i$.
The allowances $\epsilon_S$ of a flow instance are integrable, so that the total allowance $R$ is finite.

For a parent $S = U \cup \set{j}$ of a face $U$ and a label $\ell \in U \cup \set{\ast}$, the \emph{supply} of $\ell$ from $S$ to $U$ is the function
\begin{equation}\label{eq:supply}
    b^{S \to U}_\ell(x) = - F^S_{\ell, j}(x) \text{ for } x \in \Omega_U.
\end{equation}
By~\eqref{eq:facetnormals} it is the outward normal component of $F^S_\ell$ on the coordinate facet $x_j = 0$ of $\Omega_S$, positive when the flow leaves the parent toward the child, and it is determined by the parent's flow.
In this notation the conservation condition~\eqref{eq:cons} on $\Omega_U$ reads
\[
    \divergence F^U_\ell \le \sum_{j \notin U} b^{U \cup \set{j} \to U}_\ell + \epsilon_U \text{ almost everywhere on } \Omega_U
\]
for every $\ell \in U \cup \set{\ast}$, where the parent sum is empty for $U = [q]$.
This coupling between consecutive dimensions is different from normal continuity, which concerns one label flow across the cell walls within a single face.
There the two cells share a vector field whose tangential components may differ across the wall, whereas here two different vector fields on two different faces interact through one scalar.

The capacity condition~\eqref{eq:cap} bounds the difference of two label flows across coordinate subsets $I$, which are the cuts of the complete graph on the coordinate set $S$ at the point in question, whereas an interface between two label regions can have any orientation.
The next lemma converts the finitely many subset inequalities into the inequality for every normal that \cref{lem:polyhedral,lem:oneface} apply at the interface.

\begin{lemma}[Cut condition]\label{lem:cutcondition}
    Let $z \in \RR^S$ satisfy $\sum_{i \in S} z_i = 0$, let $w_{ij} = w_{ji} \ge 0$ for $i \ne j$ in $S$, and suppose that for every nonempty proper $I \subset S$,
    \[
        \abs*{\sum_{i \in I} z_i} \le \sum_{i \in I, \, j \in S \setminus I} w_{ij}.
    \]
    Then for every $\nu \in \RR^S$,
    \[
        \abs{\nu \cdot z} \le \sum_{i < j \in S} w_{ij} \abs{\nu_i - \nu_j}.
    \]
\end{lemma}
\begin{proof}
    Let $I_\theta = \set{i : \nu_i > \theta}$ and integrate over $\theta$ from $\min_i \nu_i$ to $\max_i \nu_i$.
    Since $\sum_i z_i = 0$, and a coordinate pair $i, j$ is separated by $I_\theta$ for an interval of $\theta$ of length $\abs{\nu_i - \nu_j}$,
    \begin{align*}
        \abs{\nu \cdot z}
         & = \abs*{\int \sum_{i \in I_\theta} z_i \dif \theta} \\
         & \le \int \sum_{i \in I_\theta, \, j \notin I_\theta} w_{ij} \dif \theta
         &  & \because\ \text{the subset inequalities} \\
         & = \sum_{i < j} w_{ij} \abs{\nu_i - \nu_j}. \qedhere
    \end{align*}
\end{proof}

Applied at a point $x$ of a cell with $z = F^S_\ell(x) - F^S_m(x)$ and $w_{ij} = w^S_{ij}(x)$, the lemma turns capacity~\eqref{eq:cap} into
\begin{equation}\label{eq:normalbound}
    \abs{\nu(F^S_\ell - F^S_m)} \le \sum_{i < j \in S} w^S_{ij} \abs{\nu_i - \nu_j} \text{ for every } \nu \in \RR^S,
\end{equation}
and taking $\nu_i = \one{i \in I}$ recovers~\eqref{eq:cap}, so the two forms are equivalent.
An interface with unit normal $\nu$ is crossed by the direction $e_i - e_j$ at rate $\abs{\nu_i - \nu_j}$, so the right side of~\eqref{eq:normalbound} is the cost density of the interface.
Thus, at every point an interface between any two labels, in any orientation, has cost density at least the normal difference of the two label flows across it.
The proof writes each entry as $\nu_i - \min_s \nu_s = \int \one{i \in I_\theta} \dif \theta$, so $\nu \cdot z$ is an integral over $\theta$ of the subset sums $\sum_{i \in I_\theta} z_i$ that capacity bounds.
The cost density of an interface with normal $\nu$ is the same integral of the cut weights of the sets $I_\theta$ in the complete graph on $S$.

\subsection{Summation over Faces}\label{sec:summation}

The first inequality of the chain is deterministic.
It uses only the two conditions that compare the flows across the boundaries of the faces, the terminal condition and conservation, together with the labels a non-opposite cut may use there.
For a triangle $S = \set{1, 2, 3}$ and its side $U = \set{1, 2}$, the boundary term of $\Lambda_S(C)$ includes $\int_{\Omega_U} b^{S \to U}_{C(x)}(x) \dif x$ and the flux jump $\Lambda_U(C)$ includes $- \int_{\Omega_U} \divergence F^U_{C(x)}(x) \dif x$.
The two integrals use the same points, measure, and labels, label $3$ is not allowed on the side, and conservation for the labels $1$, $2$, and $\ast$ bounds their sum below by $- \int_{\Omega_U} \epsilon_U \dif x$.

\begin{lemma}[Summation over faces]\label{lem:summation}
    Let the label flows of a weighting satisfy the terminal condition, and conservation with allowances $\epsilon_U$ and total allowance $R = \sum_{\abs{U} \ge 2} \int_{\Omega_U} \epsilon_U \dif x$.
    Then every realization $C$ of a random non-opposite cut of $\Delta_q$ satisfies
    \[
        \sum_{\abs{S} \ge 2} \Lambda_S(C) + \cost[\mathrm{term}]{C} \ge B - R.
    \]
\end{lemma}
\begin{proof}
    The boundary of $\Omega_S$ consists of the terminal facets $\Phi^S_i$, $i \in S$, and, when $\abs{S} \ge 3$, the coordinate facets $\Omega_{S \setminus \set{j}}$, $j \in S$.
    On $\Phi^S_i$, \eqref{eq:facetnormals} and the terminal condition give $\nu_{\Omega_S}(F^S_{C(y)}(y)) = g^S_i(y) \one{C(y) = i}$, and $\one{C(y) = i} + \one{C(y) \ne i} = 1$, so the selected terminal flux and the terminal cost density add to $g^S_i$ whatever the label.
    On the coordinate facet $\Omega_U$, $U = S \setminus \set{j}$, \eqref{eq:facetnormals} gives $\nu_{\Omega_S}(F^S_{C(x)}(x)) = b^{S \to U}_{C(x)}(x)$, and $C(x) \in \supp(x) \cup \set{\ast} \subseteq U \cup \set{\ast}$ because $C$ is non-opposite.
    In particular the flux of the flow of label $j$ through $x_j = 0$ is never selected.
    The same point $x$, the same label $C(x)$, and the same measure occur in the interior integral of $\Lambda_U(C)$, by the facet normalization of \cref{sec:construction}.

    Summing~\eqref{eq:Lambda} over all $S$ and grouping the coordinate-facet terms by the child $U$ on which they lie,
    \begin{equation}\label{eq:telescope}
        \sum_{\abs{S} \ge 2} \Lambda_S(C) + \cost[\mathrm{term}]{C}
        = B + \sum_{\abs{U} \ge 2} \int_{\Omega_U} \Bigl( \sum_{j \notin U} b^{U \cup \set{j} \to U}_{C(x)}(x) - \divergence F^U_{C(x)}(x) \Bigr) \dif x,
    \end{equation}
    where the parent sum is empty for $U = [q]$.
    Since $C(x)$ is an allowed label on $\Omega_U$, conservation shows that the integrand is at least $- \epsilon_U(x)$ almost everywhere, so the right side is at least $B - R$.
\end{proof}

Non-oppositeness is used exactly to ensure $C(x) \in U \cup \set{\ast}$, so that the conservation inequality of the child applies to the selected label.
The lemma is deterministic and linear in the labels, and it holds verbatim for fractional labelings.
For a cut with flat interfaces, $\Lambda_S(C)$ is the integrated normal jump across them, and the next lemma states this classical case, which the general argument reproduces in expectation.
We write $n \cdot v$ for the Euclidean inner product in the fixed independent coordinates of \cref{sec:construction}.
In the covector representation, a unit normal $n$ in the coordinates obtained by deleting $x_p$ is the vector $\nu \in \RR^S$ with $\nu_i = n_i$ for $i \ne p$ and $\nu_p = 0$, so that $n \cdot (e_i - e_j) = \nu_i - \nu_j$.

\begin{lemma}[Polyhedral cuts]\label{lem:polyhedral}
    Suppose the label flows on $\Omega_S$ have normal continuity and satisfy capacity.
    Let $C$ be a labeling of $\Omega_S$ with values in $S \cup \set{\ast}$ that is constant on the interior of each polytope of a finite subdivision of $\Omega_S$ into polytopes, each contained in a cell.
    Suppose that the labels assigned by $C$ on $\partial \Omega_S$ agree, at $\sigma$-almost every boundary point, with the label of the polytope containing that point, and that no facet between two polytopes with different labels lies in $Z_S$.
    Let $J$ be the union of the facets between polytopes with different labels, and on $J$ let $n$ be the unit normal pointing from the polytope with label $\ell^-$ into the polytope with label $\ell^+$.
    Then
    \[
        \Lambda_S(C) = \int_J n \cdot (F^S_{\ell^+} - F^S_{\ell^-}) \dif \sigma \le \cost[S]{C}.
    \]
\end{lemma}
\begin{proof}
    Apply~\eqref{eq:divthm} to $F^S_{\ell(A)}$ on each polytope $A$, where $\ell(A)$ is its label, and add.
    Facets on $\partial \Omega_S$ give the boundary term of~\eqref{eq:Lambda}, by the assumption on the boundary labels.
    On a facet shared by two polytopes with the same label the two terms cancel, since the outward normals are opposite, and if the facet is a cell facet the normal components agree across it by normal continuity.
    On a facet of $J$ the two outward contributions are $n \cdot F^S_{\ell^-}$ and $- n \cdot F^S_{\ell^+}$, which add to $- n \cdot (F^S_{\ell^+} - F^S_{\ell^-})$.
    Rearranging gives the identity.

    For the inequality, a step $h(e_i - e_j)$ crosses a flat piece of $J$ exactly from a slab of starting points of volume $\abs{h} \abs{n \cdot (e_i - e_j)}$ times the surface measure of the piece, so $\mu^S_{ij}(C) = \abs{n \cdot (e_i - e_j)} \sigma$ on $J$ and
    \[
        \cost[S]{C} = \sum_{i < j \in S} \int_J w^S_{ij} \abs{n \cdot (e_i - e_j)} \dif \sigma.
    \]
    At $\sigma$-almost every point of $J$ the two labels lie in $S \cup \set{\ast}$ and the point is interior to a cell, so capacity and \cref{lem:cutcondition} give~\eqref{eq:normalbound} there with $\nu$ the covector of $n$, and integrating over $J$ gives the inequality.
\end{proof}

We note that the inequality in \cref{lem:polyhedral} also appears in a different form in the paired calibrations of Lawlor and Morgan~\cite{LM94}, who use it to certify that a given cut is optimal.
See also \cite{Brakke91} and \cite{CCP12}.

The assumption on the boundary labels cannot simply be omitted.
The non-opposite labeling that assigns $i$ when $x_i \ge T$ and $\ast$ otherwise has no interior crossings in any truncated face and leaves every terminal edge uncut, so its cost is zero, although its boundary labels disagree with its interior labels.
It has no finite approximation factor.
We must use the approximation factor both to control the amount of interface and to recover the actual labels on the boundary.

\subsection{Crossing Measures and Regularity}\label{sec:instance}

The cost of a cut is the weight of the edges it severs.
In the continuous setting an edge in direction $v$ is severed when the label changes along a line in direction $v$, so we count these changes, ignoring changes on sets of length zero.

\begin{definition}[Crossing measures and cost]\label{def:crossing}
    For a map $c$ from an interval to a finite set, the \emph{number of changes} of $c$ is the supremum of $\#\set{s : c(t_s) \ne c(t_{s + 1})}$ over all finite increasing sequences $t_1 < \cdots < t_m$.
    The \emph{essential number of changes} $N(c)$ is the least number of changes of a map that agrees with $c$ almost everywhere.
    For a map on an open subset of $\RR$, $N$ is the sum over its connected components.
    Let $C$ be a Borel labeling of $\Omega_S$ with values in a finite set, and let $v \in V_S$ be a nonzero tangent vector.
    Parametrize the lines in direction $v$ as $t \mapsto y + t v$, with $y$ in a hyperplane $H$ transversal to $v$ and the measure $\dif y$ on $H$ for which $\dif x = \dif y \dif t$.
    For an open set $A \subseteq \intr \Omega_S$, the \emph{crossing mass} of $C$ in $A$ in direction $v$ is
    \[
        \mu_v(C)(A) = \int_H N\bigl(t \mapsto C(y + t v) \text{ on } \set{t : y + t v \in A}\bigr) \dif y.
    \]
    The labeling $C$ has \emph{finite crossings} on $\Omega_S$ if $\mu_{e_i - e_j}(C)(\intr \Omega_S) < \infty$ for all $i < j$ in $S$.
    Then each $\mu_v(C)$ extends uniquely to a finite Borel measure on $\intr \Omega_S$ (\cref{fact:slicing} below), the \emph{crossing measure} in direction $v$, and we write $\mu^S_{ij}(C) = \mu_{e_i - e_j}(C)$.
    For a weighting of $\Delta_q$ and a Borel labeling $C$ of $\Delta_q$ with values in $[q] \cup \set{\ast}$, the cost on $\Omega_S$ and the terminal cost of \cref{sec:overview-proof} are
    \[
        \cost[S]{C} = \sum_{i < j \in S} \int_{\intr \Omega_S} w^S_{ij} \dif \mu^S_{ij}(C)
        \text{ and }
        \cost[\mathrm{term}]{C} = \sum_{\abs{S} \ge 2} \sum_{i \in S} \int_{\Phi^S_i} g^S_i(y) \one{C(y) \ne i} \dif \sigma(y),
    \]
    with $\cost[S]{C} = + \infty$ if $C$ does not have finite crossings on $\Omega_S$, and the cost~\eqref{eq:cost} of $C$ is $\cost{C} = \sum_{\abs{S} \ge 2} \cost[S]{C} + \cost[\mathrm{term}]{C}$.
\end{definition}

The extension of the crossing masses to Borel measures is a statement of the theory of functions of bounded variation, which we introduce after it.

\begin{fact}[Crossing measures are derivative measures]\label{fact:slicing}
    Let $C$ be a Borel labeling of $\Omega_S$ with values in a finite set, let $u_\ell = \one{C = \ell}$ for each value $\ell$, and for a tangent vector $v \in V_S$ consider the linear functional
    \[
        T_{\ell, v}(\psi) = - \int_{\intr \Omega_S} u_\ell(x) \, \partial_v \psi(x) \dif x
    \]
    on $\psi \in C_c^1(\intr \Omega_S)$, where $\partial_v \psi$ is the directional derivative of $\psi$.
    Then $C$ has finite crossings on $\Omega_S$ if and only if, for every value $\ell$ and every tangent vector $v$, there is a finite signed Borel measure $m_{\ell, v}$ on $\intr \Omega_S$ with $T_{\ell, v}(\psi) = \int \psi \dif m_{\ell, v}$ for all $\psi \in C_c^1(\intr \Omega_S)$.
    In that case, for every nonzero $v \in V_S$ and every open $A \subseteq \intr \Omega_S$,
    \begin{equation}\label{eq:crossingBV}
        \mu_v(C)(A) = \frac{1}{2} \sum_\ell \abs{m_{\ell, v}}(A),
    \end{equation}
    where $\abs{m}$ denotes the total variation measure of a signed measure $m$.
    In particular $\mu_v(C)$ extends uniquely to a finite Borel measure on $\intr \Omega_S$.
\end{fact}

This is the slicing theory of functions of bounded variation~\cite[Sections~3.2 and~3.11]{AFP00}.
In one variable the total variation of the distributional derivative of an indicator is its essential number of changes, and in several variables the variation in direction $v$ is the integral of the variations of the one-dimensional restrictions along lines parallel to $v$.
At each change along a line exactly two indicators change, by one each, which is the factor $\frac{1}{2}$ in~\eqref{eq:crossingBV}.
For small $h$ the crossing measure $\mu^S_{ij}(C)(A)$ is approximately $\abs{h}^{-1}$ times the volume of the points $x \in A$ with $C(x) \ne C(x + h(e_i - e_j))$, the density of severed edges of length $\abs{h}$ in that direction.
\Cref{lem:variation} makes this heuristic exact in expectation.
A single point with a different label costs nothing, and the values of the weights on the cell boundaries $Z_S$ play no role, since $\mu^S_{ij}(C)(Z_S) = 0$ almost surely for the random cuts we consider (\cref{lem:variation}).
Since the weights are bounded Borel functions and the crossing measures are finite Borel measures whenever $C$ has finite crossings, the cost~\eqref{eq:cost} is defined for every Borel non-opposite cut, with the value $+ \infty$ if $C$ does not have finite crossings on some face.

Fix a face $S$, write $\Omega = \Omega_S$, and fix $d = \abs{S} - 1$ independent coordinates by deleting one coordinate.
The set $Z = Z_S$ is the union of the interior cell boundaries.
For a label $\ell \in S \cup \set{\ast}$ let $u_\ell(x) = \one{C(x) = \ell}$ be its indicator.

The functional $T_{\ell, v}$ of \cref{fact:slicing} differentiates the smooth test function $\psi$, not the bounded, possibly discontinuous $u_\ell$.
When it is represented by a finite signed measure $m_{\ell, v}$, that measure assigns a signed mass $m_{\ell, v}(A)$ to each Borel set $A$.
The function $u_\ell$ has \emph{bounded variation}, $u_\ell \in BV(\intr \Omega)$, when its derivative functionals in all coordinate directions have finite representing measures.
We collect these into the vector measure $m_\ell$, so that $m_{\ell, v} = v \cdot m_\ell$ for every direction $v$~\cite[Section~3.1]{AFP00}.
A set whose indicator has bounded variation is a \emph{set of finite perimeter}.

For a one-dimensional example, let label $1$ occupy $(-1, t)$ and label $2$ occupy $(t, 1)$, with $-1 < t < 1$.
The value at $t$ does not affect the integrals.
For direction $v = 1$ and a test function supported inside $(-1, 1)$,
\[
    T_{1, 1}(\psi) = - \int_{-1}^t \psi'(x) \dif x = - \psi(t)
    \text{ and }
    T_{2, 1}(\psi) = - \int_t^1 \psi'(x) \dif x = \psi(t).
\]
Thus, $m_{1, 1} = - \delta_t$ and $m_{2, 1} = \delta_t$, where $\delta_t$ is the unit point mass at $t$.
Both total variations are $\delta_t$, so their sum counts the same crossing twice and dividing by two counts it once, as in~\eqref{eq:crossingBV}, and integrating a weight $\varphi$ gives $\int \varphi \dif \delta_t = \varphi(t)$, the weight at the crossing.

Now let $t$ be uniform on $(-1, 1)$.
Every realization has the point mass $\delta_t$ as its crossing measure, while $\EE{\delta_t(A)}$ is half the length of $A$.
This is the form in which volume controls the crossing measures of a random cut, in expectation only, never realization by realization.
By \cref{fact:slicing}, $C$ has finite crossings on $\Omega$ if and only if every $u_\ell$ belongs to $BV(\intr \Omega)$, and then the crossing measure is $\mu_v(C) = \frac{1}{2} \sum_\ell \abs{m_{\ell, v}}$.
For $v \ne 0$ it is $\norm{v}$ times the essential number of label changes along lines parallel to $v$, integrated over the transversal directions in the fixed coordinates.

The regularity that the proof needs is not imposed on the cuts.
It is derived from the approximation factor, by the next lemma.
Its part (ii) is also the estimate that bounds the expected cost in \cref{lem:expectedcost}, and part (iii) is what lets the later arguments ignore the cell walls.

\begin{lemma}[Finite variation and expected crossings]\label{lem:variation}
    Let $C$ be a random non-opposite cut of $\Delta_q$ with approximation factor $\alpha$, and let $S$ be a face with $\abs{S} \ge 2$.
    \begin{enumerate}[(i)]
        \item Almost surely, every indicator $u_\ell = \one{C = \ell}$ belongs to $BV(\intr \Omega_S)$.
              In particular $C$ has finite crossings on $\Omega_S$.
        \item For every $i < j$ in $S$ and every bounded Borel function $\varphi \ge 0$ on $\intr \Omega_S$,
              \begin{equation}\label{eq:expectedcrossing}
                  \EE{\int \varphi \dif \mu^S_{ij}(C)} \le \alpha \int_{\Omega_S} \varphi \dif x.
              \end{equation}
        \item Consequently, almost surely $\mu^S_{ij}(C)(Z_S) = 0$ for all $i < j$ in $S$, and for the weights of any weighting
              \[
                  \EE{\cost[S]{C}} \le \alpha \sum_{i < j \in S} \int_{\Omega_S} w^S_{ij} \dif x.
              \]
    \end{enumerate}
\end{lemma}
\begin{proof}
    Write $\Omega = \Omega_S$.
    Let $\varphi \ge 0$ be a continuous, compactly supported weight on $\intr \Omega$, and define the weighted variation of $u_\ell$ in direction $v$ by
    \[
        V_v(u_\ell; \varphi) = \sup_{\substack{\psi \in C_c^1(\intr \Omega) \\ \abs{\psi} \le \varphi}} \int u_\ell \, \partial_v \psi \dif x.
    \]
    Both $\psi$ and $- \psi$ are allowed, so this is also $\sup_{\abs{\psi} \le \varphi} \abs{T_{\ell, v}(\psi)}$, and once $m_{\ell, v}$ exists, approximation of its positive and negative parts by such tests gives $V_v(u_\ell; \varphi) = \int \varphi \dif \abs{m_{\ell, v}}$.

    For sufficiently small $h > 0$, the support of $\psi$ and its translates lie inside $\intr \Omega$.
    Uniform convergence of the difference quotients of $\psi$, followed by a change of variables in the translated term, gives
    \begin{align*}
        \int u_\ell \, \partial_v \psi \dif x
         & = \lim_{h \downarrow 0} \int u_\ell(x) \frac{\psi(x) - \psi(x - h v)}{h} \dif x \\
         & = \lim_{h \downarrow 0} \int \frac{u_\ell(x) - u_\ell(x + h v)}{h} \, \psi(x) \dif x
         &  & \because\ \text{translate the second integral} \\
         & = - \lim_{h \downarrow 0} \int \psi(x) \frac{u_\ell(x + h v) - u_\ell(x)}{h} \dif x.
    \end{align*}
    Consequently, $\abs{T_{\ell, v}(\psi)} \le \liminf_{h \downarrow 0} \int \varphi(x) \abs{u_\ell(x + h v) - u_\ell(x)} h^{-1} \dif x$, and since the right side does not depend on $\psi$, the same bound holds for $V_v(u_\ell; \varphi)$.
    All limit inequalities below are taken along one fixed sequence $h \downarrow 0$.
    At any two points, either no indicator changes or exactly two change by one, so
    \[
        \frac{1}{2} \sum_\ell \abs{u_\ell(x + h v) - u_\ell(x)} = \one{C(x + h v) \ne C(x)}.
    \]

    Now take $v = e_i - e_j$, so that $d(x, x + h v) = h$.
    Only the labeling $C$ is random.
    The set $\Omega$, the direction $v$, the weight $\varphi$, and the tests are fixed.
    The difference-quotient integrands are nonnegative, which permits Fatou's lemma and Tonelli's exchange of expectation and spatial integration:
    \begin{align}
        \EE{\frac{1}{2} \sum_\ell V_v(u_\ell; \varphi)}
         & \le \EE{\liminf_{h \downarrow 0} \frac{1}{2} \sum_\ell \int \varphi(x) \frac{\abs{u_\ell(x + h v) - u_\ell(x)}}{h} \dif x}
         &  & \because\ \text{the pathwise bounds} \notag \\
         & \le \liminf_{h \downarrow 0} \EE{\frac{1}{2} \sum_\ell \int \varphi(x) \frac{\abs{u_\ell(x + h v) - u_\ell(x)}}{h} \dif x}
         &  & \because\ \text{Fatou} \notag \\
         & = \liminf_{h \downarrow 0} \frac{1}{h} \int \varphi(x) \Pr{C(x + h v) \ne C(x)} \dif x
         &  & \because\ \text{Tonelli and two indicators} \notag \\
         & \le \alpha \int_\Omega \varphi \dif x
         &  & \because\ \text{the approximation factor}.
        \label{eq:expectedvariation}
    \end{align}
    The supremum was bounded for each realization before taking the expectation.
    No exchange of expectation and supremum is used.

    Choose continuous compactly supported weights $0 \le \varphi_n \uparrow 1$, each equal to one on a compact set, with those sets exhausting the interior.
    Every compactly supported continuous weight between zero and one is bounded by some $\varphi_n$, so monotone convergence and~\eqref{eq:expectedvariation} give
    \[
        \EE{\frac{1}{2} \sum_\ell \sup_{\substack{\varphi \in C_c(\intr \Omega) \\ 0 \le \varphi \le 1}} V_v(u_\ell; \varphi)} = \lim_{n \to \infty} \EE{\frac{1}{2} \sum_\ell V_v(u_\ell; \varphi_n)} \le \alpha \int_\Omega 1 \dif x < \infty.
    \]
    A nonnegative quantity with finite expectation is finite almost surely.
    There are finitely many labels, faces, and directions $e_i - e_j$, so all these variations are finite on one event of probability one.

    On that event each functional satisfies $\abs{T_{\ell, v}(\psi)} \le \norm{\psi}_\infty \sup_{0 \le \varphi \le 1} V_v(u_\ell; \varphi)$, which extends $T_{\ell, v}$ uniquely by uniform approximation from $C_c^1(\intr \Omega)$ to the continuous functions vanishing at the boundary.
    The Riesz representation theorem~\cite[Theorem~1.54]{AFP00} then supplies a unique finite signed measure $m_{\ell, v}$ with $T_{\ell, v}(\psi) = \int \psi \dif m_{\ell, v}$.
    It is this boundedness, together with linearity, that permits the theorem.
    The conclusion is a measure, which may have point masses.
    The directions $e_i - e_j$ span $V_S$, so applying this to a spanning set gives the vector measure $m_\ell$ and $u_\ell \in BV(\intr \Omega)$, which is (i).
    By \cref{fact:slicing}, $C$ has finite crossings on $\Omega$.

    On the same event, $V_v(u_\ell; \varphi) = \int \varphi \dif \abs{m_{\ell, v}}$ and~\eqref{eq:crossingBV} turn~\eqref{eq:expectedvariation} into $\EE{\int \varphi \dif \mu^S_{ij}(C)} \le \alpha \int_\Omega \varphi \dif x$ for continuous compactly supported $\varphi \ge 0$.
    For an open set $A \subseteq \intr \Omega$, increase such weights to $\one{A}$ to obtain $\EE{\mu^S_{ij}(C)(A)} \le \alpha \int_A 1 \dif x$.
    For a Borel set $A$, contain it in open sets whose volumes decrease to its volume, and monotonicity gives the same bound.
    Finally, approximate a bounded Borel $\varphi \ge 0$ by increasing simple functions, which proves~\eqref{eq:expectedcrossing}.

    For (iii), the set $Z_S$ is fixed and has volume zero, so $\varphi = \one{Z_S}$ in~\eqref{eq:expectedcrossing} gives $0 \le \EE{\mu^S_{ij}(C)(Z_S)} \le 0$, and the nonnegative mass $\mu^S_{ij}(C)(Z_S)$ vanishes almost surely.
    The weights $w^S_{ij}$ are bounded Borel functions, even where they jump on cell boundaries, so $\varphi = w^S_{ij}$ in~\eqref{eq:expectedcrossing}, summed over $i < j$, bounds $\EE{\cost[S]{C}}$ as stated.

    \emph{Measurability.}
    Exhaust the interior by a countable family of compact sets and, for each, choose a countable dense subset of its admissible tests in the $C^1$ norm.
    The integral is continuous in that norm, so the same supremum is obtained from the countable union, whose individual evaluations are measurable.
    The approximations by open and Borel sets above, followed by the usual monotone-class extension, also make the measure evaluations on Borel sets measurable in the realization.
\end{proof}

The bound~\eqref{eq:expectedcrossing} concerns expectations only, as the uniform-$t$ example shows.
The crossing measure of an individual cut lives on its interfaces and is not bounded by volume, but a fixed set of volume zero carries no crossing mass almost surely.

The capacity condition bounds the normal difference of one pair of label flows by the combined directional crossing cost for that same normal.
To use it we need the derivative measures of all the indicators to be carried by one common interface with exactly two neighboring labels and one normal at almost every point, so that the flux and every directional crossing measure refer to the same geometry.
The structure theory of partitions into sets of finite perimeter provides precisely this representation.
The following is an adapted consequence of De Giorgi's structure theorem~\cite[Theorem~3.59]{AFP00} and the local structure of Caccioppoli partitions~\cite[Theorem~4.17]{AFP00}.

\begin{fact}[Two-label structure of a finite-variation cut]\label{fact:structure}
    Fix the open interior $\intr \Omega$ in the chosen $d$ independent coordinates, and a Borel labeling $C \colon \intr \Omega \to S \cup \set{\ast}$ whose every indicator $u_\ell = \one{C = \ell}$ belongs to $BV(\intr \Omega)$, with derivative measures $m_\ell$.
    There exist a Borel set $J \subseteq \intr \Omega$, the \emph{interface}, with finite surface measure $\sigma(J)$, measurable distinct neighboring labels $\ell^+, \ell^-$ on $J$, and a measurable unit normal $n$ on $J$ pointing toward $\ell^+$, with the following properties.
    Except for a set of surface measure zero, $J$ is covered by countably many continuously differentiable hypersurfaces.
    At $\sigma$-almost every $x \in J$, exactly the two indicated labels occupy the two sides, in the sense that
    \[
        \lim_{r \downarrow 0} \int_{B(0, 1)} \abs{u_{\ell^\pm(x)}(x + r z) - \one{\pm z \cdot n(x) > 0}} \dif z = 0,
    \]
    the ball and the normal being taken in the fixed coordinates, while every other label's indicator tends to zero in the same sense.
    For every label $\ell$ and every Borel $A \subseteq \intr \Omega$,
    \begin{equation}\label{eq:structure}
        m_\ell(A) = \int_{J \cap A} \bigl( \one{\ell = \ell^+} - \one{\ell = \ell^-} \bigr) n \dif \sigma.
    \end{equation}
    Equivalently, every compactly supported continuously differentiable vector field $G$ on $\intr \Omega$ satisfies
    \[
        \int_{\intr \Omega} u_\ell \, \divergence G \dif x = - \int_J \bigl( \one{\ell = \ell^+} - \one{\ell = \ell^-} \bigr) n \cdot G \dif \sigma.
    \]
\end{fact}

De Giorgi's theorem supplies the surface representation and the half-space limits for each set of finite perimeter, and the partition theorem says that almost every point of these surfaces belongs to exactly two label boundaries, with opposite normals.
Their union, minus the exceptional null set, is $J$.
This is the measure-theoretic interface carrying the derivative measures, not the entire topological boundary of the label regions, and the integral over $J \cap A$ in~\eqref{eq:structure} counts only the part of $A$ lying on $J$.
At an interface point the two nonzero signed contributions in~\eqref{eq:structure} are $n$ for $\ell^+$ and $- n$ for $\ell^-$.
Pairing them with their label flows gives $n \cdot F_{\ell^+} - n \cdot F_{\ell^-}$, whereas taking directional absolute values and dividing by two gives $\frac{1}{2}(\abs{n \cdot v} + \abs{- n \cdot v}) = \abs{n \cdot v}$.
Thus, for every Borel $A \subseteq \intr \Omega$, every tangent vector $v$, and bounded Borel vector fields $F_\ell$,
\begin{equation}\label{eq:interfacecrossing}
    \begin{aligned}
        \mu_v(C)(A) & = \int_{J \cap A} \abs{n \cdot v} \dif \sigma, \\
        \sum_\ell \int_A F_\ell \cdot \dif m_\ell & = \int_{J \cap A} n \cdot (F_{\ell^+} - F_{\ell^-}) \dif \sigma,
    \end{aligned}
\end{equation}
where the vector-measure integral in the second line is the sum of the integrals of each component of the field against the corresponding coordinate measure.

These are the cost and flux descriptions of the same interface, since all directions see the same two labels and the same normal, and the capacity condition compares the two integrands pointwise.
We combine~\eqref{eq:interfacecrossing} with \cref{lem:variation}(iii).
Since the directions $e_i - e_j$ span, the continuous function $\sum_{i < j} \abs{n \cdot (e_i - e_j)}$ has a positive minimum $c_S$ over unit normals, so almost surely
\[
    \sigma(J \cap Z_S) \le c_S^{-1} \sum_{i < j} \mu^S_{ij}(C)(Z_S) = 0
    \text{ and }
    \EE{\sigma(J)} \le c_S^{-1} \sum_{i < j} \EE{\mu^S_{ij}(C)(\intr \Omega_S)} < \infty,
\]
and by~\eqref{eq:structure} also $\abs{m_\ell}(Z_S) = 0$, where $\abs{m_\ell}$ is the total variation of the vector measure.

\subsection{Recovering the Boundary Labels}\label{sec:boundary}

\Cref{fact:structure} concerns interfaces in the open interior and tests only compactly supported vector fields.
It does not identify the labels that $C$ assigns on the outer boundary $\partial \Omega_S$, which are the labels used in the flux jump~\eqref{eq:Lambda} and in the cancellation between faces of \cref{lem:summation}.
The counterexample after \cref{lem:polyhedral} shows that interior information alone cannot determine them.
The next identity supplies the connection, in expectation, using the approximation factor.
Integration by parts first yields an interior expression, and the approximation factor makes the probability of disagreement between a boundary point and a nearby interior point tend to zero at a controlled rate.
A normalized collar integral then recovers the selected flux with the actual boundary labels.

\begin{lemma}[Boundary-flux identity]\label{lem:boundaryflux}
    Let $C$ be a random non-opposite cut of $\Delta_q$ with approximation factor $\alpha$, let $S$ be a face with $\abs{S} \ge 2$, and let the label flows $F^S_\ell$ on $\Omega_S$ have normal continuity.
    Almost surely $C$ has finite crossings on $\Omega_S$, so that \cref{fact:structure} provides an interface $J$ with labels $\ell^\pm$ and normal $n$.
    With this notation,
    \begin{equation}\label{eq:expectedflux}
        \EE{\Lambda_S(C)} = \EE{\int_J n \cdot (F^S_{\ell^+} - F^S_{\ell^-}) \dif \sigma}.
    \end{equation}
\end{lemma}
\begin{proof}
    Write $\Omega = \Omega_S$, $F_\ell = F^S_\ell$, $Z = Z_S$, and $F_C(x) = F_{C(x)}(x)$.
    The interface exists almost surely by \cref{lem:variation}(i), and $\sigma(J \cap Z) = 0$ and $\abs{m_\ell}(Z) = 0$ almost surely by the remark after~\eqref{eq:interfacecrossing}.
    Work in the fixed coordinates of \cref{sec:instance}.

    \emph{Cutoff.}
    For each facet $f$ of $\Omega$ let $a_f(x)$ be the inward distance to its hyperplane, and let $s(x) = \min_f a_f(x)$.
    Choose the piecewise affine cutoff $\chi_\delta = \min(1, \max(0, s/\delta - 1))$, which is zero within distance $\delta$ of the boundary and one beyond $2 \delta$.
    Define the modified fields $G_\ell = \chi_\delta F_\ell$, whose product rule on each smooth piece is $\divergence G_\ell = \chi_\delta \divergence F_\ell + F_\ell \cdot \nabla \chi_\delta$.
    Refine the cells along the finitely many planes where the cutoff changes formula.
    Across new facets $G_\ell$ is continuous, and across old facets its normal components agree by normal continuity and the continuity of $\chi_\delta$.
    Apply~\eqref{eq:divthm} to $\psi G_\ell$ on each piece, for $\psi \in C_c^1(\intr \Omega)$.
    Shared boundary terms cancel and outer boundary terms vanish, giving
    \[
        \int G_\ell \cdot \nabla \psi \dif x = - \int \psi \bigl( \chi_\delta \divergence F_\ell + F_\ell \cdot \nabla \chi_\delta \bigr) \dif x.
    \]
    Thus, the displayed bounded, cellwise function is also the distributional divergence of $G_\ell$, with no extra measure on the cell facets.

    \emph{Mollification.}
    Choose a nonnegative smooth $\rho$ supported in the unit ball with $\int \rho \dif x = 1$, let $\rho_\eta(z) = \eta^{-d} \rho(z/\eta)$, extend $G_\ell$ by zero outside $\Omega$, and let $G_{\ell, \eta}(x) = \int \rho_\eta(z) G_\ell(x - z) \dif z$.
    For $0 < \eta < \delta/2$ the averaging enlarges the support by at most $\eta$, so $G_{\ell, \eta}$ is a compactly supported smooth field inside $\intr \Omega$, and the integral identity of \cref{fact:structure} gives
    \[
        \int u_\ell \, \divergence G_{\ell, \eta} \dif x = - \int G_{\ell, \eta} \cdot \dif m_\ell.
    \]
    The divergence of $G_{\ell, \eta}$ is the same weighted average of $\divergence G_\ell$, by the preceding integration identity, and converges to it in $L^1$ as $\eta \downarrow 0$.
    Since $0 \le u_\ell \le 1$, the left side converges to $\int u_\ell \divergence G_\ell \dif x$.
    Outside the cell walls $Z$ the field $G_\ell$ is continuous, so $G_{\ell, \eta}(x) \to G_\ell(x)$ there.
    Moreover, $\norm{G_{\ell, \eta}(x)} \le \sup_x \norm{G_\ell(x)}$, and $\abs{m_\ell}$ is finite with $\abs{m_\ell}(Z) = 0$, so dominated convergence gives the limit of the right side.
    Summing over labels and using~\eqref{eq:interfacecrossing},
    \begin{equation}\label{eq:GG}
        \int \chi_\delta \divergence F_C \dif x + \int F_C \cdot \nabla \chi_\delta \dif x = - \int_J \chi_\delta \, n \cdot (F_{\ell^+} - F_{\ell^-}) \dif \sigma.
    \end{equation}

    \emph{Limits.}
    As $\delta \downarrow 0$ the cutoff increases to one in the interior.
    The first term converges in expectation to $\EE{\int_\Omega \divergence F_C \dif x}$, since the divergences are bounded and $\Omega$ has finite volume.
    The right side converges to $- \EE{\int_J n \cdot (F_{\ell^+} - F_{\ell^-}) \dif \sigma}$, dominated by $2 \sup_{\ell, x} \norm{F_\ell(x)} \sigma(J)$, which has finite expectation.
    It remains to show that the negative of the middle term converges in expectation to the selected outward boundary flux.

    \emph{Collar.}
    Where $f$ uniquely minimizes $a_f$ and $\delta < a_f < 2 \delta$, we have $- \nabla \chi_\delta = n_f/\delta$, with $n_f$ the outward unit normal of $f$.
    Write such a point as $x = y - t n_f$, with $y \in f$ and $\delta < t < 2 \delta$.
    The volume element is $\dif \sigma(y) \dif t$.
    Exclude from $f$ a strip of surface measure $O(\delta)$ around its edges and around the intersections of cell facets with $f$.
    On the remaining set $K_\delta$ the segment $[y, y - 2 \delta n_f]$ lies in one cell and the nearest outer facet is $f$.
    This follows from the finitely many affine inequalities defining the polytopes, since away from the intersection strips the distances to all other facet hyperplanes exceed a fixed multiple of $\delta$.
    Conversely, every point of the collar assigned to $f$ projects to a point of $f$, so the excluded part contributes only $O(\delta)$ after the $1/\delta$ normalization.

    Inside a cell let $K$ be a common Lipschitz bound for the label flows and $M$ a bound for their norms, and let $c_f$ be one half of the $\ell_1$ norm of $n_f$ lifted to the tangent space $V_S$, so that $d(y - t n_f, y) = c_f t$.
    For $y \in K_\delta$ the approximation factor gives $\Pr{C(y - t n_f) \ne C(y)} \le \alpha \, d(y - t n_f, y) = c_f \alpha t$.
    There are two errors, moving from $y - t n_f$ to $y$ within the same label flow, and using different label flows when the labels disagree.
    Adding and subtracting the flow with label $C(y - t n_f)$ evaluated at $y$, the first error is at most $K t$ by the Lipschitz bound, and the second is zero for equal labels and at most $2 M$ otherwise, so
    \[
        \EE{\abs{\nu_f(F_{C(y - t n_f)}(y - t n_f)) - \nu_f(F_{C(y)}(y))}} \le K t + 2 M \Pr{C(y - t n_f) \ne C(y)} \le (K + 2 M c_f \alpha) t.
    \]
    Averaging over the collar,
    \[
        \frac{1}{\delta} \int_\delta^{2 \delta} \int_{K_\delta} \EE{\abs{\nu_f(F_{C(y - t n_f)}(y - t n_f)) - \nu_f(F_{C(y)}(y))}} \dif \sigma(y) \dif t
        \le \frac{\sigma(f) (K + 2 M c_f \alpha)}{\delta} \int_\delta^{2 \delta} t \dif t = O(\delta).
    \]
    Together with the discarded strips, this proves
    \[
        - \EE{\int_\Omega F_C \cdot \nabla \chi_\delta \dif x} \longrightarrow \EE{\int_{\partial \Omega} \nu_\Omega(F_{C(y)}(y)) \dif \sigma(y)}.
    \]
    In dimension one, facets are endpoints, $\sigma$ is counting measure, and no strip is discarded for small $\delta$.
    The same estimate applies.
    Taking expectations and limits in~\eqref{eq:GG} proves~\eqref{eq:expectedflux}.
\end{proof}

The three hypotheses play separate roles.
Normal continuity cancels the artificial flux that the cell walls would otherwise contribute to the distributional divergence.
The zero crossing mass on the walls, \cref{lem:variation}(iii), permits the piecewise argument and the passage to the limit against $m_\ell$.
The approximation factor connects the interior limit to the labels that $C$ assigns on the boundary, which are the labels the cancellation between faces needs.
The boundary integral in~\eqref{eq:expectedflux} deliberately uses those labels.

\subsection{One Face and Expected Cost}\label{sec:completing}

The second inequality of the chain compares, face by face, the selected flux with the cost.

\begin{lemma}[One face]\label{lem:oneface}
    Suppose the label flows on $\Omega_S$ have normal continuity and satisfy capacity for the weights of a weighting.
    Then every random non-opposite cut $C$ of $\Delta_q$ with a finite approximation factor satisfies
    \[
        \EE{\Lambda_S(C)} \le \EE{\cost[S]{C}}.
    \]
\end{lemma}
\begin{proof}
    On $\Omega_S$ the labels of a non-opposite cut lie in $S \cup \set{\ast}$.
    Almost surely $C$ has an interface $J$ with $\sigma(J \cap Z_S) = 0$, by \cref{lem:variation} and the remark after~\eqref{eq:interfacecrossing}.
    At a point of $J \setminus Z_S$ we are in the interior of a cell, the labels $\ell^\pm$ are allowed, and capacity holds, so \cref{lem:cutcondition} with $z = F^S_{\ell^+} - F^S_{\ell^-}$ and $\nu$ the covector of $n$ gives~\eqref{eq:normalbound} there.
    Hence,
    \begin{align*}
        \EE{\Lambda_S(C)}
         & = \EE{\int_J n \cdot (F^S_{\ell^+} - F^S_{\ell^-}) \dif \sigma}
         &  & \because\ \text{\cref{lem:boundaryflux}} \\
         & \le \EE{\int_J \sum_{i < j \in S} w^S_{ij} \abs{n \cdot (e_i - e_j)} \dif \sigma}
         &  & \because\ \text{capacity and \cref{lem:cutcondition}} \\
         & = \EE{\sum_{i < j \in S} \int_{\intr \Omega_S} w^S_{ij} \dif \mu^S_{ij}(C)}
         &  & \because\ \eqref{eq:interfacecrossing} \\
         & = \EE{\cost[S]{C}}
         &  & \because\ \text{\cref{def:crossing}}.
    \end{align*}
    Weights and flows may have different values on cell facets, but those facets carry no crossing mass and no interface, so the pointwise comparison on $J \setminus Z_S$ suffices.
\end{proof}

The third inequality of the chain uses the approximation factor alone, through \cref{lem:variation}(ii) for the interior edges and directly for the terminal edges.

\begin{lemma}[Expected cost]\label{lem:expectedcost}
    Let a weighting have fractional cost $L$.
    Every random non-opposite cut $C$ of $\Delta_q$ with approximation factor $\alpha$ satisfies $\EE{\cost{C}} \le \alpha L$.
\end{lemma}
\begin{proof}
    For each $S$, \cref{lem:variation}(iii) gives $\EE{\cost[S]{C}} \le \alpha \sum_{i < j} \int_{\Omega_S} w^S_{ij} \dif x$.
    For $y \in \Phi^S_i$ we have $d(y, e_i) = 1 - T$ and $C(e_i) = i$, so $\Pr{C(y) \ne i} \le \alpha (1 - T)$, and Fubini's theorem gives $\EE{\cost[\mathrm{term}]{C}} \le \alpha (1 - T) B$.
    Adding over the faces and comparing with~\eqref{eq:L} gives the claim.
\end{proof}

With the three lemmas proved, the proof of \cref{thm:general} in \cref{sec:overview-proof} is complete.

The bound of \cref{thm:general} can only improve when $q$ grows.

\begin{proposition}[Zero extension]\label{prop:monotone}
    A flow instance on $\Delta_q$ extends to a flow instance on $\Delta_{q'}$, for every $q' \ge q$, with the same $B$, $R$, and $L$.
\end{proposition}
\begin{proof}
    For $S \subseteq [q']$ with $S \not\subseteq [q]$, take the single cell $\calP_S = \set{\Omega_S}$ and set $w^S_{ij} = 0$, $g^S_i = 0$, $F^S_\ell = 0$, and $\epsilon_S = 0$.
    Normal continuity, capacity, and the terminal condition hold trivially on the new faces.
    In conservation for $U \subseteq [q]$, the new parents $U \cup \set{j}$ with $j \notin [q]$ contribute zero supply, so the parent sum is unchanged, and for $U \not\subseteq [q]$ both sides vanish.
    The quantities $B$, $R$, and $L$ receive only zero contributions.
\end{proof}

\Cref{sec:explicit,sec:recipe,sec:recorded} apply the theorem to the two flow instances of \cref{sec:overview-instances}.

\section{Affine Flow Instances}\label{sec:explicit}

We show how to extend the flow instance of \cref{sec:triangle} into a family of instances for every $q$ and every $T$.
We call these flows \emph{affine}.
Every truncated face is a single cell, the label flows point toward the terminal from a point that moves affinely with $x$, as on the hexagon, the weights are constant, and the only choice is one scale per dimension.
We force the scales so that the flux leaving every face through its coordinate facets equals the divergence of its children, so that an allowance is needed only on the top face, which has no parent.
The current section formally defines this affine family, and proves \cref{thm:explicit}.

\begin{theorem}[Affine flow instances]\label{lem:explicit}
    Fix $q \ge 2$ and $T \in (1/2, 1)$, and for $2 \le r \le q$ let
    \begin{equation}\label{eq:explicitscales}
        a_r = \frac{(r T - 1) (r - 2)! \, (q - r)!}{T^{r - 2} \, q!}.
    \end{equation}
    For every $S \subseteq [q]$ with $r = \abs{S} \ge 2$, let $\Omega_S$ be a single cell and choose
    \begin{equation}\label{eq:explicitfields}
        \begin{gathered}
            p^S_s(x) = \frac{T - x_s}{r T - 1} \text{ for } s \in S, \ \ F^S_i(x) = a_r \bigl( e_i - p^S(x) \bigr) \text{ for } i \in S, \text{ and } F^S_\ast = 0, \\
            w^S_{st} = \frac{a_r}{r - 1} \text{ for } s < t \in S \text{ and } g^S_i = a_r \text{ for } i \in S,
        \end{gathered}
    \end{equation}
    with allowances $\epsilon_U = 0$ for $U \ne [q]$ and $\epsilon_{[q]} = (q - 1) a_q/(q T - 1)$.
    The flows as defined above give a flow instance on $\Delta_q$, such that
    \begin{equation}\label{eq:explicitvalues}
        \begin{gathered}
            \boxed{B = \sum_{r = 2}^q \frac{(r T - 1) z^{r - 2}}{(r - 1)!} = T \Bigl( 1 - \frac{z^{q - 1}}{(q - 1)!} \Bigr)}, \ \
            \boxed{R = \frac{1 - q (1 - T)^{q - 1}}{q! \, T^{q - 2}}}, \\
            \boxed{B - R = T - \frac{1}{q! \, T^{q - 2}}}, \ \ \boxed{L = (1 - T) B + \sum_{r = 2}^q \frac{(r T - 1) \bigl( 1 - r (1 - T)^{r - 1} \bigr)}{2 (r - 1) \, (r - 1)! \, T^{r - 2}}},
        \end{gathered}
    \end{equation}
    where $z = (1 - T)/T$.
\end{theorem}

Since $x_s \le T$ on $\Omega_S$ and $\sum_s x_s = 1$, the point $p^S(x)$ lies in $\Delta_S$, and $p^S_i = 0$ exactly on the terminal facet $x_i = T$.
So the flow of label $i$ is $a_r$ times the vector from the normalized margin point $p^S(x)$ to the vertex $e_i$, as on the triangle.
At $q = 3$ and $T = 2/3$, multiplying all weights, terminal densities, label flows, and allowances by $4$ gives exactly the flow instance of \cref{prop:triangle}.

\begin{proof}
    \emph{Normal continuity} holds trivially, since each face is one cell.

    \emph{Capacity.}
    Let $I \subset S$ be nonempty and proper, and let $P_I = \sum_{s \in I} p^S_s \in [0, 1]$.
    Then $\sum_{s \in I} F^S_{\ell, s} = a_r (\one{\ell \in I} - P_I)$ for $\ell \in S$ and $0$ for $\ell = \ast$.
    Over the labels $\ell \in I$, $\ell \in S \setminus I$, and $\ast$, these values are $a_r (1 - P_I)$, $- a_r P_I$, and $0$, so they span an interval of length exactly $a_r$.
    The weight crossing $I$ is $\abs{I} (r - \abs{I}) a_r/(r - 1) \ge a_r$, so~\eqref{eq:cap} holds.

    \emph{Terminal condition.}
    On $\Phi^S_i$ we have $p^S_i = 0$, so $F^S_{\ell, i} = a_r \one{\ell = i} = g^S_i \one{\ell = i}$ for every label $\ell$.

    \emph{Conservation.}
    In every tangent direction $v$ the derivative of $p^S$ is $- v/(r T - 1)$, so the derivative of $F^S_i$ is $a_r v/(r T - 1)$, and since the tangent space has dimension $r - 1$,
    \[
        \divergence F^S_i = \frac{(r - 1) a_r}{r T - 1} \text{ for } i \in S.
    \]
    Let $U$ have $r$ coordinates and let $S = U \cup \set{j}$ be one of its $q - r$ parents.
    On $\Omega_U$, which is the facet $x_j = 0$ of $\Omega_S$, the supply~\eqref{eq:supply} is $b^{S \to U}_\ell = - F^S_{\ell, j} = a_{r + 1} p^S_j = a_{r + 1} T/((r + 1) T - 1)$ for every $\ell \in U$, and $0$ for $\ell = \ast$.
    The scales~\eqref{eq:explicitscales} satisfy
    \[
        (q - r) \, a_{r + 1} \, \frac{T}{(r + 1) T - 1} = \frac{(r - 1) a_r}{r T - 1} \text{ for } 2 \le r < q,
    \]
    since both sides equal $(r - 1)! \, (q - r)!/(q! \, T^{r - 2})$.
    So on $\Omega_U$ the divergence of every allowed label's flow equals the supply arriving from its parents, and~\eqref{eq:cons} holds with zero allowance for every $U \ne [q]$.
    For the label $\ast$ both sides are zero.
    The top face $\Omega_{[q]}$ has no parent, and its allowance $(q - 1) a_q/(q T - 1)$ is its whole divergence.

    \emph{The values.}
    There are $\binom{q}{r}$ faces with $r$ coordinates.
    Each has volume $(1 - r (1 - T)^{r - 1})/(r - 1)!$, the simplex minus its $r$ disjoint corners.
    Each has $\binom{r}{2}$ directions, each of weight $a_r/(r - 1)$, and $r$ terminal facets, each of measure $(1 - T)^{r - 2}/(r - 2)!$.
    Multiplying out and substituting~\eqref{eq:explicitscales} gives the sums for $B$ and $L$, and $R$ is the top-face divergence $(q - 1) a_q/(q T - 1)$ times the volume of the top face.
    The sum for $B$ telescopes, because its $r$th term is $T \bigl( z^{r - 2}/(r - 2)! - z^{r - 1}/(r - 1)! \bigr)$, and the formula for $B - R$ follows.
\end{proof}

\begin{proof}[Proof of \cref{thm:explicit}]
    Take $q = 10$ and $T = 3/5$ in \cref{lem:explicit}.
    Substitution in~\eqref{eq:explicitvalues} yields
    \[
        B = \frac{55801301}{93002175}, \ \ R = \frac{43289}{2645395200}, \text{ and } L = \frac{8180229294583}{16665989760000},
    \]
    whose ratio $(B - R)/L$ is $9999320418500/8180229294583 = 1.2223765445\ldots$.
\end{proof}

By \cref{thm:general}, every random non-opposite cut of $\Delta_{10}$ with approximation factor $\alpha$ satisfies $\alpha \ge 1.2223765445$, and $\lim_{k \to \infty} G_k \ge 1.2223765445$.

For $T = 3/5$, the dependence on the number of coordinates is
\begin{center}
    \begin{tabular}{c*{8}{c}}
        \toprule
        $q$ & $3$ & $4$ & $5$ & $6$ & $7$ & $8$ & $9$ & $10$ \\
        \midrule
        $(B - R)/L$ & $1.0985$ & $1.1664$ & $1.2019$ & $1.2162$ & $1.2208$ & $1.2221$ & $1.2223$ & $1.2224$ \\
        \bottomrule
    \end{tabular}
\end{center}

The computer-aided flow instance of \cref{sec:recorded} improves the affine construction by allowing piecewise quadratic label flows and weights on many cells, and by choosing them by linear programming, which \cref{sec:recipe} describes next.

\section{Finding and Verifying a Flow Instance by Linear Programming}\label{sec:recipe}

We find the flow instance behind \cref{thm:main} by computer search, which chooses the weighting and the label flows that bound its cuts together.
Note that this is different from solving the relaxation~\eqref{eq:ckr}, where the weights are fixed and the embedding is optimized.
This section describes the family of flow instances in which we search, the linear program that finds the best one, and the finite list of integer inequalities that proves that the one found is a flow instance.
\Cref{sec:recorded} then states the instance and completes the proof of \cref{thm:main}.

We impose two simplifying restrictions on the family we search in.
The first is symmetry: we only consider flow instances that are mapped to themselves when the coordinates of $\Delta_q$ are permuted and the labels are renamed in the same way.
Such an instance is determined by its restriction to one region per dimension, the part of one face on which the coordinates are sorted, so instead of a weighting and label flows on every face we choose them on $q - 1$ regions.
The second is finiteness: each region is subdivided into simplices, the cells of \cref{sec:construction}, and on each cell every weight and every component of every label flow is a polynomial of degree at most two.
A flow instance in the family is then a finite list of coefficients, and then $B$, $R$, and $L$ are linear.

The conditions of \cref{def:certificate} are linear in the weighting, the label flows, and the allowances, and they are homogeneous, so the flow instance of largest value in the family is found by a linear program.
However, there is one challenge.
The conditions must hold at every point of every cell, and at each point they are linear in the coefficients, but there are infinitely many points.
We write the polynomials in the Bernstein basis, whose basis functions are nonnegative, so that an inequality between coefficients implies the same inequality at every point.
This replaces every pointwise condition by finitely many linear inequalities, at the price of searching a smaller family.

\subsection{Symmetric Flow Instances}\label{sec:symmetry}

Let $r$ be the number of coordinates of a face, so that $\Omega_S$ with $\abs{S} = r$ has dimension $r - 1$, and $2 \le r \le q$.
All faces with $r$ coordinates are copies of one another, and inside one face the $r!$ orderings of its coordinates are mirror images of one another.
We take one representative for all of them.
For local coordinates on a face with $r$ coordinates use the indices $0, \ldots, r - 1$, and let
\[
    \calO_r = \set{x \in \Delta_r : x_0 \ge x_1 \ge \cdots \ge x_{r - 1}, \ x_0 \le T} \text{ for } 2 \le r \le q
\]
be the \emph{ordered region}.
It is the part of a truncated face with $r$ coordinates on which the coordinates are sorted.
Its boundary consists of the terminal facet $x_0 = T$, the \emph{equality facets} $x_j = x_{j + 1}$ for $0 \le j \le r - 2$, and, for $r \ge 3$, the coordinate facet $x_{r - 1} = 0$, which is $\calO_{r - 1}$ with a zero coordinate appended.

For a face $S \subseteq [q]$ with $\abs{S} = r$ and a bijection $\pi \colon \set{0, \ldots, r - 1} \to S$, the \emph{copy} $\calO_\pi = \set{y \in \Omega_S : y_{\pi(0)} \ge \cdots \ge y_{\pi(r - 1)}}$ is the image of $\calO_r$ under the coordinate permutation $x \mapsto y$ with $y_{\pi(s)} = x_s$.
The copies with $r$ coordinates cover the faces $\Omega_S$ with $\abs{S} = r$ up to the sets where two coordinates are equal, which have volume zero, and there are
\[
    M_r = \binom{q}{r} r! = \frac{q!}{(q - r)!}
\]
of them, one for each face with $r$ coordinates and each order of its coordinates.
A function $f$ on $\calO_r$ is \emph{transported} to $\calO_\pi$ as $y \mapsto f(x)$, a label flow $F_i$ for a local label $i$ is transported to the flow of the label $\pi(i)$ with components $F_{\pi(i), \pi(s)}(y) = F_{i, s}(x)$, a weight $w_{st}$ to $w_{\pi(s) \pi(t)}$, a terminal density $g_0$ on $x_0 = T$ to $g_{\pi(0)}$, and the label $\ast$ is fixed.

A \emph{symmetric} flow instance is specified, \emph{unscaled}, on each $\calO_r$, and its weighting, label flows, and allowances on every copy are the transported ones \emph{divided by} $M_r$.
The division is a convention.
It makes $B$, $R$, and $L$ the plain integrals over the $q - 1$ ordered regions, and it makes conservation on $\calO_r$ an inequality with a single supply term.
Every face with $r$ coordinates is the coordinate facet $x_j = 0$ of $q - r$ faces with $r + 1$ coordinates, and
\begin{equation}\label{eq:parentcount}
    M_{r + 1} = (q - r) M_r, \text{ or equivalently } \frac{q - r}{M_{r + 1}} = \frac{1}{M_r}.
\end{equation}
For $q = 4$ and $r = 2$ this reads $M_2 = 12$ and $M_3 = 24$.
An interval has two parent triangles, and if each supplies an unscaled quantity divided by $24$, the interval receives that quantity divided by $12$, its own scale.

\begin{lemma}[Coordinate symmetry]\label{lem:symmetry}
    For $2 \le r \le q$, let $\calP^{(r)}$ be a finite subdivision of $\calO_r$ into cells whose restriction to the facet $x_{r - 1} = 0$ of $\calO_r$ is the subdivision $\calP^{(r - 1)}$ of $\calO_{r - 1}$ with a zero coordinate appended.
    Let $w^{(r)}_{st} \ge 0$, $g^{(r)}_0 \ge 0$ on $x_0 = T$, label flows $F^{(r)}_i$ for $i \in \set{0, \ldots, r - 1, \ast}$ with $F^{(r)}_\ast = 0$, and $\epsilon^{(r)} \ge 0$ be quantities on $\calO_r$ of the kinds required of a weighting, label flows, and allowances in \cref{sec:overview-simplex,sec:construction}.
    Transport the cells and these quantities to every copy, dividing the weights, terminal densities, label flows, and allowances by $M_r$.
    Then the following hold.
    \begin{enumerate}[(i)]
        \item The total terminal weight, the total allowance, and the fractional cost of the transported weighting are the unscaled sums
              \[
                  B = \sum_{r = 2}^q \int_{\set{x_0 = T} \cap \calO_r} g^{(r)}_0 \dif \sigma, \ \
                  R = \sum_{r = 2}^q \int_{\calO_r} \epsilon^{(r)} \dif x, \text{ and }
                  L = \sum_{r = 2}^q \sum_{s < t} \int_{\calO_r} w^{(r)}_{st} \dif x + (1 - T) B.
              \]
        \item For $r < q$ and almost every $x \in \calO_r$, the parent sum of conservation at the transported point of any copy of $\calO_r$ equals $1/M_r$ times the unscaled supply $- F^{(r + 1)}_{\ell, r}(x, 0)$ from $\calO_{r + 1}$ across its facet $x_r = 0$, for every label $\ell \in \set{0, \ldots, r - 1, \ast}$.
        \item The transported quantities satisfy capacity and the terminal condition on every face if and only if the unscaled ones satisfy them on every $\calO_r$, and they satisfy conservation if and only if, for every $r$ and every $\ell \in \set{0, \ldots, r - 1, \ast}$, almost everywhere on $\calO_r$,
              \[
                  \divergence F^{(r)}_\ell(x) \le - F^{(r + 1)}_{\ell, r}(x, 0) + \epsilon^{(r)}(x),
              \]
              with the supply term omitted for $r = q$.
    \end{enumerate}
\end{lemma}
\begin{proof}
    Transport preserves volume and surface measure, and the copies with $r$ coordinates are $M_r$ in number and cover the faces with $r$ coordinates up to volume zero, so the integral over all copies of a transported function divided by $M_r$ is the integral of the unscaled function over $\calO_r$.
    This is (i).
    For (ii), let $y$ be the transported point of $x$ in a copy $\calO_\pi$ on a face $U$ with $\abs{U} = r$, with all coordinates of $x$ distinct and positive.
    Each parent $S = U \cup \set{j}$ contains $y$ in exactly one copy of $\calO_{r + 1}$, the one ordering $S$ by $\pi$ followed by $j$, since $y_j = 0$ is its smallest coordinate.
    On that copy the supply~\eqref{eq:supply} of the transported flow is $- F^{(r + 1)}_{\ell, r}(x, 0)/M_{r + 1}$, and summing over the $q - r$ parents and using~\eqref{eq:parentcount} gives (ii).
    For (iii), capacity and the terminal condition are pointwise inequalities and equalities that are homogeneous in the quantities and covariant under the coordinate permutation.
    The divergence of a transported flow is the transported divergence divided by $M_r$ and the allowance is divided by $M_r$, so by (ii) the inequality~\eqref{eq:cons} at $y$ is $1/M_r$ times the displayed inequality at $x$.
\end{proof}

\subsection{The Cells and the Quadratic Flows}\label{sec:subdivision}

In this section $d = r - 1$ is the dimension of a face with $r$ coordinates and never denotes the distance.
For $2 \le j \le r$ let
\[
    u_j = (\underbrace{1/j, \ldots, 1/j}_{j}, 0, \ldots, 0)
    \text{ and }
    v_j = (T, \underbrace{(1 - T)/(j - 1), \ldots, (1 - T)/(j - 1)}_{j - 1}, 0, \ldots, 0).
\]
Since $T > 1/2$, these points are ordered, and they are the vertices of $\calO_r$.
An initial subdivision is
\begin{equation}\label{eq:initialmesh}
    \conv(u_2, \ldots, u_s, v_s, \ldots, v_r) \text{ for } 2 \le s \le r.
\end{equation}
For $r = 2$ the region is the interval $[u_2, v_2]$.
For $r \ge 3$, the facets of $\calO_r$ not containing $v_r$ are the coordinate facet $\calO_{r - 1}$, on $x_{r - 1} = 0$, and the equality facet $\conv(u_2, \ldots, u_r)$, on $x_0 = x_1$.
Joining $v_r$ to these two facets and applying induction proves~\eqref{eq:initialmesh}.

Fix integers $n \ge 1$ and $r_0 \ge 2$.
For $2 \le r \le r_0$, subdivide each simplex in~\eqref{eq:initialmesh} into $n^{r - 1}$ simplices as follows.
Keep its displayed vertex order.
Take an integer vector $z$ with $0 \le z_1 \le \cdots \le z_d \le n - 1$ and a permutation $\sigma$ of $[d]$, and form the path $z, \ z + e_{\sigma(1)}, \ \ldots, \ z + (1, \ldots, 1)$, keeping it only when every vector along it is weakly increasing.
For each path vertex $y$, use the barycentric coefficients
\begin{equation}\label{eq:subdivision}
    \frac{1}{n} (n - y_d, \ y_1, \ y_2 - y_1, \ \ldots, \ y_d - y_{d - 1})
\end{equation}
in the ordered original simplex.
Each path gives one small simplex.
Cumulative barycentric coordinates identify the original simplex with $\set{0 \le y_1 \le \cdots \le y_d \le n}$, and dividing its unit cubes along coordinate orders shows that this is a subdivision.
Its restriction to a facet uses the same rule with the inherited vertex order.

For $r > r_0$, append a zero coordinate to the subdivision of $\calO_{r - 1}$ and join each small simplex to $v_r$.
For every simplex $f$ of the subdivision of $\calO_{r - 1}$ that lies on $x_0 = x_1$, also include $\conv(f, u_r, v_r)$.
The two-facet decomposition above shows that these simplices cover $\calO_r$ with disjoint interiors.
In both cases, the subdivision of the coordinate facet of $\calO_r$ is exactly the one already chosen for $\calO_{r - 1}$, as \cref{lem:symmetry} requires, and the cells of \cref{sec:construction}, the subdivision $\calP_S$ of every face $\Omega_S$, are the copies of these small simplices.
For $r = 3$ the region $\calO_3$ is the quadrilateral with vertices $u_2, v_2, v_3, u_3$, the initial mesh consists of the triangles $u_2 v_2 v_3$ and $u_2 u_3 v_3$, and each is cut into $n^2$ cells, $9$ for the recorded $n = 3$.
The volumes give an exact check of the geometry:
\begin{equation}\label{eq:volumes}
    \abs{\calO_r} = \frac{1 - r (1 - T)^{r - 1}}{r! \, (r - 1)!}
    \text{ and }
    \abs{\set{x_0 = T} \cap \calO_r} = \frac{(1 - T)^{r - 2}}{(r - 2)! \, (r - 1)!}.
\end{equation}
The first subtracts the $r$ disjoint corner simplices $\set{x_i \ge T}$ from $\Delta_r$ and divides by $r!$.
The second divides the terminal facet into the $(r - 1)!$ orders of its remaining coordinates.

Let $P$ be a cell with ordered vertices $a_0, \ldots, a_d$, volume $V_P$, and barycentric coordinates $\lambda_0, \ldots, \lambda_d$.
On $P$ every component of every label flow and every weight is a quadratic polynomial, which we write in the \emph{Bernstein basis}
\begin{equation}\label{eq:quadraticbasis}
    F(x) = \sum_s F_{ss} \lambda_s(x)^2 + 2 \sum_{s < t} F_{st} \lambda_s(x) \lambda_t(x),
\end{equation}
with one coefficient $F_{st}$ for each pair $s \le t$ of cell vertices.
The mixed coefficient $F_{st}$ is generally not the value at the midpoint of $a_s a_t$.
Three properties of the basis carry the rest of the section.
The basis functions are nonnegative and sum to one, so an inequality between the coefficients of two quadratics implies the same inequality between their values at every point of the cell.
On a facet of $P$, the basis functions of the pairs in the facet restrict to the Bernstein basis of the facet and the others vanish, so a parent's flux through a facet and a child's quantities on that facet are compared in one basis.
Every basis function has integral $2 V_P/(r (r + 1))$, so an integral over a cell is a sum of coefficients.

The first property is sufficient, not necessary.
On $[0, 1]$ the polynomial $(1 - 2 t)^2 = (1 - t)^2 - 2 t (1 - t) + t^2$ is nonnegative, but its coefficients $1, - 1, 1$ in this basis are not.
We impose every pointwise condition of \cref{def:certificate} coefficient by coefficient.
This restricts the flow instances we search for and verify, while the resulting flow instance still covers every cut in \cref{thm:general}.

\subsection{The Linear Program}\label{sec:search}

All conditions are homogeneous, so the value $(B - R)/L$ does not change when all weights, terminal densities, label flows, and allowances are multiplied by the same positive number.
With fixed cells, the search for the best flow instance in the family is therefore the following problem, linear in the unknown functions $w$, $g$, $F$, and $\epsilon$:
\[
    \begin{array}{ll}
        \text{maximize}   & B - R \\
        \text{subject to} & L = 1,\ \ w^S_{ij} \ge 0,\ \ g^S_i \ge 0,\ \ \epsilon_U \ge 0,\ \ \textstyle\sum_i F^S_{\ell, i} = 0,\ \ F^S_\ast = 0, \\
                          & \text{normal continuity, capacity, the terminal condition, and conservation.}
    \end{array}
\]
By \cref{lem:symmetry} the unknowns are the coefficients on the ordered regions, and conservation couples $\calO_r$ only to $\calO_{r + 1}$.
The supplies in conservation are derived from the label flows and are not separate unknowns.
Normal continuity and the terminal condition are equalities between coefficients on facets, and \cref{sec:reconstruction} chooses the unknowns so that they hold identically.
Capacity and conservation are inequalities between coefficients, imposed through the first property of the basis, with both signs for each absolute value.
The objective and the normalization are sums of coefficients by the third property.
The result is a finite linear program, and \cref{sec:recorded} describes how it was solved.

\subsection{Representing a Flow by Its Boundary Fluxes}\label{sec:reconstruction}

Normal continuity, the terminal condition, and the supply to a child all concern the normal flux of a label flow through a facet of a cell.
We therefore take the fluxes as the unknowns and recover the flow from them.
A facet shared by two cells then carries one coefficient, read with opposite orientations by the two cells, so normal continuity holds by construction.
The terminal condition becomes a rule on the coefficients of the terminal facet, and the supply to the child is the coefficient on the facet $x_{r - 1} = 0$.
The remaining freedom of a flow inside a cell, which changes it without changing its normal components, is a second set of unknowns.

Orient each facet $f$ outward from its first incident cell in a fixed order of the cells.
Write $s_{P, f} = 1$ for that cell and $s_{P, f} = - 1$ for the other incident cell, if any, and let $o_{P, f}$ be the vertex of $P$ opposite $f$.
The table consists of three arrays of integers.
They are integers $H_{i, f, st}$ for each label $i$, facet $f$, and pair of vertices $a_s, a_t$ of $f$, including $s = t$ and symmetric in $s, t$.
They are integers $Z_{i, P, st}$ for each label $i$, cell $P$, and cell edge $a_s a_t$ with $s < t$.
They are nonnegative integers $W_{P, st, ij}$ for each cell $P$, quadratic coefficient $st$, and direction $e_i - e_j$ with $i < j$.

Let $h_{i, f, st} = H_{i, f, st}/D$ and $z_{i, P, st} = Z_{i, P, st}/D$ for a fixed positive integer $D$.
Here $h_{i, f, st}$ is the prescribed quadratic coefficient of the normal component of $F_i$ on $f$, multiplied by the measure of the entire facet, in the chosen orientation.
It is a facet-integrated coefficient, to be distinguished from the pointwise supply $b$.
The vector coefficients of the flows and the weight coefficients are
\begin{align}
    V_P F_{i, P, st} & = \frac{1}{d} \sum_{\substack{f \subset \partial P \\ a_s, a_t \in f}} s_{P, f} \, h_{i, f, st} \Bigl( \frac{a_s + a_t}{2} - o_{P, f} \Bigr) + \begin{cases} z_{i, P, st} (a_s - a_t), & s < t, \\ 0, & s = t, \end{cases} \label{eq:reconstruction} \\
    w_{ij, P, st} & = \frac{W_{P, st, ij}}{D V_P}, \label{eq:weights}
\end{align}
extended inside $P$ by~\eqref{eq:quadraticbasis}, with $F_\ast = 0$.
Nonnegative coefficients give nonnegative weights.
The values of the weights on cell boundaries are irrelevant, since those boundaries have volume zero and carry no crossing measure almost surely (\cref{lem:variation}).

Two \emph{boundary rules} are imposed on $H$.
On an equality facet $x_j = x_{j + 1}$ of $\calO_r$, the normal coefficients for the labels $j$ and $j + 1$ are negatives of one another, and those of all other labels vanish.
On the terminal facet $x_0 = T$, the coefficients for label $0$ are nonnegative and those of all other labels vanish.

\begin{lemma}[Reconstruction]\label{lem:reconstruction}
    The flows~\eqref{eq:reconstruction} are tangent to their face, and on every facet $f$ of a cell $P$ the quadratic normal component of $F_i$, multiplied by the measure of $f$, has coefficients exactly $s_{P, f} h_{i, f, st}$ for $a_s, a_t \in f$ and zero for the other pairs.
    In particular the flows have normal continuity inside $\calO_r$.
    If $H$ satisfies the equality-facet rule, the transported flows have normal continuity across the copies of $\calO_r$.
    If $H$ satisfies the terminal-facet rule, they satisfy the terminal condition with $g_0$ the normal density of $F_0$ on $x_0 = T$, which is nonnegative.
\end{lemma}
\begin{proof}
    Every vector $(a_s + a_t)/2 - o_{P, f}$ and $a_s - a_t$ is a difference of points of the face, hence tangent.
    The outward normal covector of $P$ integrated over the facet opposite a vertex $o$ is $- d V_P \nabla \lambda_o$.
    Applied to $(a_s + a_t)/2 - o$ it gives $d V_P$ whenever $a_s, a_t$ lie in that facet, and its value on the corresponding vector for any other facet containing $a_s, a_t$ is zero.
    The term $a_s - a_t$ has zero normal component on every facet containing both vertices.
    Hence the normal-component coefficients of $F_i$ on $f$, multiplied by the measure of $f$, are exactly $s_{P, f} h_{i, f, st}$, while on $f$ the basis functions of pairs not contained in $f$ vanish.

    Since the two cells sharing a facet use the same coefficients with opposite orientations, their quadratic normal components agree, which is normal continuity inside $\calO_r$.
    Two copies of $\calO_r$ meeting along $x_j = x_{j + 1}$ are related by the transposition of the coordinates $j$ and $j + 1$ and of the labels $j$ and $j + 1$.
    The equality-facet rule says that this transposition matches the normal components of all labels across the facet, which is normal continuity across copies.

    On $x_0 = T$, the outward normal component is the $0$th component by~\eqref{eq:facetnormals}, and the terminal-facet rule says that it vanishes for every label other than $0$ and is a nonnegative quadratic for label $0$, which is the terminal condition with that quadratic as $g_0$.
    Finally, after prescribing the normal coefficients of a mixed pair, the remaining freedom is exactly the direction $a_s - a_t$, so no vector coefficients are omitted by~\eqref{eq:reconstruction}.
\end{proof}

\subsection{The Finite Conditions}\label{sec:conditions}

With the flows and weights given by the table, two conditions of \cref{def:certificate} remain, and we reduce each to a statement about the integers.
Capacity becomes a list of integer inequalities, which the table must satisfy.
Conservation is satisfied by definition, because the allowance is defined as whatever the divergence exceeds the supply by, and it is paid for in $R$.

\paragraphname{Capacity.}
Since $\max_\ell y_\ell - \min_\ell y_\ell \le c$ if and only if $\abs{y_\ell - y_m} \le c$ for all $\ell, m$, capacity at a coefficient is equivalent to requiring, for every nonempty proper $I \subset \set{0, \ldots, r - 1}$,
\begin{equation}\label{eq:capmaxmin}
    \max_{\ell \in \set{0, \ldots, r - 1, \ast}} \sum_{i \in I} F_{\ell, P, st, i} - \min_{\ell \in \set{0, \ldots, r - 1, \ast}} \sum_{i \in I} F_{\ell, P, st, i} \le \sum_{i \in I, \, j \notin I} w_{ij, P, st},
\end{equation}
where $F_{\ast, P, st} = 0$.
Only subsets containing coordinate $0$ are needed, since complementary subsets give the same condition.
For each $r$, write the vertex coordinates as $p_{s, j}/Q_r$ with a common positive integer denominator $Q_r$, and let
\[
    N_{i, j}(P, st) = \sum_{f \supset \set{a_s, a_t}} s_{P, f} H_{i, f, st} (p_{s, j} + p_{t, j} - 2 p_{o_{P, f}, j}) + \begin{cases} 2 d Z_{i, P, st} (p_{s, j} - p_{t, j}), & s < t, \\ 0, & s = t, \end{cases}
\]
so that $V_P F_{i, P, st, j} = N_{i, j}(P, st)/(2 d Q_r D)$ by~\eqref{eq:reconstruction}, with $N_{\ast, j} = 0$.

\begin{lemma}[Integer capacity]\label{lem:intcap}
    Suppose that for every $r$, every cell $P$ of $\calO_r$, every quadratic coefficient $st$, and every nonempty proper $I \subset \set{0, \ldots, r - 1}$,
    \begin{equation}\label{eq:integercapacity}
        \max_{\ell \in \set{0, \ldots, r - 1, \ast}} \sum_{j \in I} N_{\ell, j}(P, st) - \min_{\ell \in \set{0, \ldots, r - 1, \ast}} \sum_{j \in I} N_{\ell, j}(P, st) \le 2 d Q_r \sum_{i \in I, \, j \notin I} W_{P, st, ij}.
    \end{equation}
    Then the flows~\eqref{eq:reconstruction} and the weights~\eqref{eq:weights} satisfy capacity at every point of every cell of every $\calO_r$.
\end{lemma}
\begin{proof}
    Multiplying~\eqref{eq:capmaxmin} by $2 d Q_r D V_P$ gives~\eqref{eq:integercapacity}, so the hypothesis is capacity between corresponding coefficients, for every pair of labels and every subset.
    For a fixed pair of labels and a fixed $I$, the inequality is linear in the flow and weight coefficients, so the first property of the basis transfers it to every point of the cell.
\end{proof}

\paragraphname{Conservation.}
The divergence of a quadratic flow is affine, and the residual is computed from the flows and their parent fluxes and then bounded above by a nonnegative polynomial.
The sign is forced by $\divergence F \le b + \epsilon$, so the allowance coefficients are bounded below by the shortfalls, not above.
Let $A_{P, i, s} = V_P \divergence F_{i, P}(a_s)$.
Differentiating~\eqref{eq:quadraticbasis} and substituting~\eqref{eq:reconstruction} gives
\begin{equation}\label{eq:integrateddiv}
    A_{P, i, s} = \frac{1}{d} \sum_{f \ni a_s} s_{P, f} \Bigl( 2 h_{i, f, ss} + \sum_{a_t \in f \setminus \set{a_s}} h_{i, f, st} \Bigr) + 2 \Bigl( \sum_{t < s} z_{i, P, ts} - \sum_{t > s} z_{i, P, st} \Bigr).
\end{equation}
An affine function, written in the quadratic basis, has coefficient at $st$ equal to the average of its values at $a_s$ and $a_t$.
For $r < q$, let $f^+$ be the coordinate facet of $\calO_{r + 1}$ that matches $P$ after a zero coordinate is appended.
Its coefficient $h_{i, f^+, st}$ is the coefficient of the unscaled supply from $\calO_{r + 1}$, multiplied by $V_P$.
Define
\begin{equation}\label{eq:residual}
    e_{P, st} = \max \Bigl( 0, \ \max_{0 \le i < r} \Bigl\{ \frac{A_{P, i, s} + A_{P, i, t}}{2} - h_{i, f^+, st} \Bigr\} \Bigr),
\end{equation}
with the parent coefficient replaced by zero when $r = q$.

\begin{lemma}[Integer conservation]\label{lem:intcons}
    Let $\epsilon^{(r)}$ be the quadratic on each cell $P$ of $\calO_r$ with coefficients $e_{P, st}/V_P$.
    Then $\epsilon^{(r)} \ge 0$, and the transported flows and allowances satisfy conservation on every face of $\Delta_q$.
\end{lemma}
\begin{proof}
    The coefficients $e_{P, st}$ are nonnegative, so $\epsilon^{(r)} \ge 0$ by the nonnegative basis.
    By~\eqref{eq:integrateddiv} and the averaging rule, $(A_{P, i, s} + A_{P, i, t})/(2 V_P)$ is the coefficient at $st$ of $\divergence F_{i, P}$.
    By \cref{lem:reconstruction} applied to $f^+$ and~\eqref{eq:facetnormals}, $h_{i, f^+, st}/V_P$ is the coefficient of the unscaled supply $- F^{(r + 1)}_{i, r}(\cdot, 0)$ on $P$.
    So~\eqref{eq:residual} says that $\epsilon^{(r)}$ dominates, coefficient by coefficient, the difference between each label's divergence and its unscaled supply, and the nonnegative basis transfers this to every point.
    For $\ast$ both are zero.
    This is the inequality of \cref{lem:symmetry}(iii), which gives conservation on every face after transport and scaling.
\end{proof}

\paragraphname{The quantities $B$, $R$, and $L$.}
A facet has $r (r - 1)/2$ quadratic coefficients and a cell has $r (r + 1)/2$, each with the same integral, so by \cref{lem:symmetry}(i) and~\eqref{eq:weights}
\begin{gather}
    B = \sum_{r = 2}^{q} \sum_{f \text{ on } x_0 = T} \frac{2}{r (r - 1)} \sum_{s \le t : \, a_s, a_t \in f} h_{0, f, st}
    \text{ and }
    R = \sum_{r = 2}^{q} \sum_{P \subset \calO_r} \frac{2}{r (r + 1)} \sum_{s \le t} e_{P, st}, \label{eq:BRLtable} \\
    L = \sum_{r = 2}^{q} \sum_{P \subset \calO_r} \frac{2}{r (r + 1) D} \sum_{s \le t} \sum_{i < j} W_{P, st, ij} + (1 - T) B.
    \label{eq:Ltable}
\end{gather}

\begin{proposition}[Soundness of the finite checks]\label{prop:recipe}
    Fix $q \ge 2$, a rational $T \in (1/2, 1)$, integers $n \ge 1$, $r_0 \ge 2$, and $D \ge 1$, and build the subdivision of \cref{sec:subdivision}.
    Let integers $H$, $Z$, and $W \ge 0$ satisfy the two boundary rules of \cref{sec:reconstruction} and the integer capacity inequality~\eqref{eq:integercapacity} for every cell, coefficient, and subset.
    Then the flows~\eqref{eq:reconstruction}, the weights~\eqref{eq:weights}, the terminal densities given by the normal density of $F_0$ on $x_0 = T$, and the allowances with coefficients~\eqref{eq:residual}, transported to every face by \cref{lem:symmetry}, form a flow instance on $\Delta_q$ whose $B$, $R$, and $L$ are given by~\eqref{eq:BRLtable} and~\eqref{eq:Ltable}.
    In particular, if $L > 0$, then every random non-opposite cut of $\Delta_q$ with approximation factor $\alpha$ satisfies $\alpha \ge (B - R)/L$, and $\lim_k G_k \ge (B - R)/L$.
\end{proposition}
\begin{proof}
    The quantities are of the kinds required in \cref{sec:overview-simplex,sec:construction}, since the cells are simplices, the weights and flows are quadratics on each cell, and the terminal densities are nonnegative quadratics.
    Normal continuity holds inside each ordered region and across ordered regions by \cref{lem:reconstruction}, and the terminal condition by the same lemma.
    Capacity holds on every $\calO_r$ by \cref{lem:intcap}, hence on every face by \cref{lem:symmetry}(iii), and conservation holds by \cref{lem:intcons}.
    The formulas for $B$, $R$, and $L$ integrate the basis functions, by \cref{lem:symmetry}(i).
    The last assertion is \cref{thm:general}.
\end{proof}

The validity of a flow instance given by a table therefore depends only on exact integer checks, not on how the table was found.

\subsection{The Flow Instance on $\Delta_{11}$}\label{sec:recorded}

\begin{proof}[Proof of \cref{prop:certificate}]
    The flow instance uses $q = 11$, $T = 249/400$, $n = 3$, $r_0 = 5$, and $D = 230000000000$ in the subdivision of \cref{sec:subdivision}, with quadratic flows and weights on every cell.
    Its ordered regions contain $2{,}937$ cells:
    \begin{center}
        \begin{tabular}{c*{10}{r}}
            \toprule
            $r$ & $2$ & $3$ & $4$ & $5$ & $6$ & $7$ & $8$ & $9$ & $10$ & $11$ \\
            \midrule
            Cells & $3$ & $18$ & $81$ & $324$ & $351$ & $378$ & $405$ & $432$ & $459$ & $486$ \\
            \bottomrule
        \end{tabular}
    \end{center}
    The table consists of the integers $H$, $Z$, and $W$ of \cref{sec:reconstruction} for $r = 2, \ldots, 11$.
    It satisfies the two boundary rules and the integer capacity inequalities~\eqref{eq:integercapacity} for every cell, coefficient, and subset, which we checked in integer and rational arithmetic, together with the geometry of the subdivision, by the checker described at the end of this section.
    By \cref{prop:recipe} the table defines a flow instance on $\Delta_{11}$ with $T = 249/400$, and the sums~\eqref{eq:BRLtable} and~\eqref{eq:Ltable}, evaluated in exact arithmetic, are
    \begin{equation}\label{eq:recorded-values}
        B = \frac{1725000000167}{1725000000000}, \ \
        R = \frac{2593033}{910800000000000}, \text{ and }
        L = \frac{43107286191734959}{53130000000000000}.
    \end{equation}
    Their ratio is
    \[
        \frac{B - R}{L} = \gapvalue = 1.232506254686709004\ldots. \qedhere
    \]
\end{proof}

\begin{proof}[Proof of \cref{thm:main}]
    By \cref{prop:certificate} there is a flow instance on $\Delta_{11}$ with $L > 0$ and value $\beta = \gapvalue$, so by \cref{thm:general} every random non-opposite cut of $\Delta_{11}$ with approximation factor $\alpha$ satisfies $\alpha \ge \beta$.
    \Cref{lem:reduction} with $q = 11$ and this $\beta$ gives $\lim_{k \to \infty} G_k \ge \beta > 1.2325$ and, for every $\eta > 0$ and all sufficiently large $k$, a finite $k$-terminal instance whose integrality gap is at least $\beta - \eta$.
\end{proof}

\paragraphname{How the flow instance was found.}
The linear program of \cref{sec:search} was solved with the allowance set to zero, so that the normalization is $B = 1$ and $L$ is minimized, and the allowance entered only after rounding.
To keep the program small, the label flows were restricted in the higher dimensions.
For $r \le 5$ every label has its own quadratic flow and the weights are free.
For $6 \le r \le 10$ all labels share one quadratic flow, the flow of label $i$ adds to it the constant vector $A (e_i - u_r)$ for one amplitude $A$ per dimension, where $u_r$ is the barycenter, and every weight equals $A/(r - 1)$, as in the affine family of \cref{sec:explicit}.
For $r = 11$ the shared flow is affine.
These restrictions describe the search, not the flow lower bound, and we make no claim of optimality within the family.
The program was solved by an interior-point method in floating point arithmetic.
Its solution was rounded to the common denominator $D$, the allowance was set to the residual~\eqref{eq:residual} of the rounded table, and the weights were rounded upward and increased where capacity required it, then reduced wherever every inequality remained valid.
The proof of \cref{prop:certificate} uses the rounded table alone.

For these weights, the lower bound on the cut cost is already within $R$ of the cost $B$ of the cut that labels every point by $\ast$, with $R/L < 3.6 \times 10^{-9}$, so improving the flow estimate for these weights can add very little to the ratio.
A substantial improvement needs different weights or geometry, or additional restrictions on the labels of one common random cut.
The affine flow instances show the mechanism in its simplest form, and the computer-aided flow instance shows how far piecewise quadratic flows on a fixed subdivision carry it.

\paragraphname{Availability.}
The table, the checker, and the search and rounding code are available at \url{https://github.com/dimitarch/multiway-cut-integrality-gap}.
The checker \path{verify.py} rebuilds the subdivision of \cref{sec:subdivision} from the stored vertices, checks the two boundary rules and the inequality~\eqref{eq:integercapacity} for every cell, coefficient, and subset in integer arithmetic, and prints $B$, $R$, $L$, and their ratio as the fractions of~\eqref{eq:recorded-values}, in about ten seconds and without any optimization software.
The floating solution of the linear program is stored as well, and the rounding and trimming steps described above are deterministic, so the table can be regenerated from it.

\paragraph{Generative AI Statement.}
The authors came up the idea of establishing a lower bound for the LP relaxation of the Minimum Multiway Cut problem using flows in a continuous setting and non-opposite cuts.
The proofs were developed through interactive sessions with GPT-6 Astra.
The authors substantially revised the proofs proposed by the LLM, identified a number of gaps and fixed them.
Claude Fable 5.1 and Opus 5.5 were used in preparing the numerical optimization code.
We take full responsibility for the contents of this paper.

\bibliographystyle{alpha}
\bibliography{bibliography}

\newpage\clearpage
\appendix
\section{Proof of the Non-Opposite Reduction}\label{app:reduction}

\Cref{lem:reduction} turns a lower bound on the approximation factor of every random non-opposite cut of $\Delta_q$ into a lower bound on $\lim_{k \to \infty} G_k$.
Its proof runs the finite-grid reduction of~\cite{AMM17}, sketched in \cref{sec:reduction}, at $\alpha = \lim_{k \to \infty} G_k$, and then extends the resulting cut from the rational points of $\Delta_q$ to a measurable cut of all of $\Delta_q$.

\begin{proof}[Proof of \cref{lem:reduction}]
    Let $\alpha = \lim_{k \to \infty} G_k$.
    The limit exists because $G_k$ is nondecreasing, it is finite because $G_k \le 3/2$ for every $k$ by the rounding of~\cite{CKR00} quoted in \cref{sec:intro}, and $G_k \le \alpha$ for every $k$.
    We construct a random non-opposite cut of $\Delta_q$ with approximation factor $\alpha$.

    Let $\Delta_{q, n}$ be the grid of points of $\Delta_q$ whose coordinates are multiples of $1/n$.
    Restrict a random cut of $\Delta_k$ with approximation factor $G_k$ to the face of $q$ terminals chosen uniformly at random, identified with $\Delta_q$, and replace every label not allowed at a point by $\ast$.
    After symmetrizing the cut over permutations of the terminals, a label outside the support of a point $x$ is equally likely to be any of the $k - \abs{\supp(x)}$ other terminals, and only $q - \abs{\supp(x)}$ of them were chosen, so the replacement changes the label at $x$ with probability at most $q/k$.
    Two points of $\Delta_{q, n}$ are at distance at least $1/n$, so the replacement adds at most $O(q n/k)$ to the approximation factor on $\Delta_{q, n}$, which vanishes as $k \to \infty$ with $n$ fixed.
    The compactness of the finite simplex of laws on the labelings of $\Delta_{q, n}$ then gives, for every $n$, a random non-opposite cut of $\Delta_{q, n}$ with approximation factor $\alpha$.
    If the bound is stated only for neighboring grid points, shortest grid paths and the union bound give it for every pair.
    Take the grids with denominators $n!$, which are nested, and a diagonal subsequence of their joint laws.
    The labels take finitely many values, so the subsequence converges on every finite set of rational points.
    The limiting finite-dimensional distributions are consistent, and the extension theorem~\cite[Section~36]{Billingsley95} gives a random cut of the rational points of $\Delta_q$ with these distributions.
    It is non-opposite and has approximation factor $\alpha$, because both properties are conditions on finitely many points.

    To extend it measurably, round every coordinate of $x$ except its first positive coordinate down to a multiple of $2^{-m}$, and assign the remaining mass to that first coordinate.
    The resulting map $\pi_m$ is Borel, has countable range, satisfies $\supp(\pi_m(x)) \subseteq \supp(x)$ and $\pi_m(e_i) = e_i$, and satisfies $d(\pi_m(x), x) \le (q - 1) 2^{-m}$.
    Thus
    \[
        \Pr{C(\pi_m(x)) \ne C(\pi_{m + 1}(x))} \le \alpha (q - 1) (2^{-m} + 2^{-m - 1}),
    \]
    which is summable in $m$, so for each fixed $x$ the labels $C(\pi_m(x))$ are eventually constant almost surely by the Borel--Cantelli lemma.

    Define $C(x)$ to be this eventual value, and the first label in $\supp(x)$ on the exceptional null event.
    The result is measurable for the product $\sigma$-algebra, since each map $(\xi, x) \mapsto C(\pi_m(x))$ is.
    It is non-opposite for every realization, because every $C(\pi_m(x))$ lies in $\supp(x) \cup \set{\ast}$, and it fixes the terminals.
    For each fixed pair $x, y$, bounded convergence gives
    \[
        \Pr{C(x) \ne C(y)} = \lim_{m \to \infty} \Pr{C(\pi_m(x)) \ne C(\pi_m(y))} \le \lim_{m \to \infty} \alpha \, d(\pi_m(x), \pi_m(y)) = \alpha \, d(x, y).
    \]
    So $C$ is a random non-opposite cut of $\Delta_q$ with approximation factor $\alpha$, and the hypothesis of the lemma gives $\alpha \ge \beta$, that is, $\lim_{k \to \infty} G_k \ge \beta$.
    For the second clause, fix $\eta > 0$.
    For all sufficiently large $k$ we have $G_k \ge \beta - \eta/2$, and since $G_k$ is a supremum over finite $k$-terminal instances, some finite $k$-terminal instance has integrality gap at least $\beta - \eta$.
\end{proof}

\section{Flow Instances on a Graph}\label{sec:graphflows}

\Cref{sec:overview-graph} defined flow instances on a finite graph $H$ and stated the flow lower bound on a graph, \cref{lem:discrete}, and its obstruction, \cref{prop:discretelimit}.
This section proves both.
A small example shows how a vertex where a label is forbidden acts as a source of that label's flow, at no cost.
The obstruction is what the construction on the simplex gets around, as \cref{sec:overview-simplex} explains.

\begin{proof}[Proof of \cref{lem:discrete}]
    Let the label flows $\varphi_\ell$ and the allowances $\epsilon_v$ form a flow instance on $H = (V, E)$, and let $C$ be a non-opposite cut, so that $C(v) \in A(v)$ for every vertex $v$.
    The net inflow of label $\ell$ at $v$ is $\operatorname{in}_\ell(v) = \sum_{u : uv \in E} \varphi_\ell(u \to v)$, where entering flow counts positively and leaving flow negatively.
    We set $g_{v, i} = 0$ for $i \notin \supp(v)$ and $g_{v, \ast} = 0$, so that $g_{v, C(v)} = \sum_{i \in [q]} g_{v, i} \one{C(v) = i}$.

    Sum the selected inflows $\operatorname{in}_{C(v)}(v)$ over the vertices.
    An edge $uv$ appears in the inflow at $v$ as $\varphi_{C(v)}(u \to v)$ and in the inflow at $u$ as $\varphi_{C(u)}(v \to u) = - \varphi_{C(u)}(u \to v)$, so it contributes $\varphi_{C(v)}(u \to v) - \varphi_{C(u)}(u \to v)$.
    This is zero if $C(u) = C(v)$ and at most $w_{uv}$ otherwise, by capacity, since $C(u), C(v) \in A(u) \cup A(v)$.
    Conservation at every vertex for its label $C(v) \in A(v)$ gives $\operatorname{in}_{C(v)}(v) \ge g_{v, C(v)} - \epsilon_v$.
    Hence
    \begin{align*}
        \sum_{uv \in E} w_{uv} \one{C(u) \ne C(v)}
         & \ge \sum_{uv \in E} \bigl( \varphi_{C(v)}(u \to v) - \varphi_{C(u)}(u \to v) \bigr)
         &                                                                                   & \because\ \text{capacity on every edge}       \\
         & = \sum_{v \in V} \operatorname{in}_{C(v)}(v)
         &                                                                                   & \because\ \text{regroup by vertex}            \\
         & \ge \sum_{v \in V} g_{v, C(v)} - R
         &                                                                                   & \because\ \text{conservation at every vertex}.
    \end{align*}
    The last sum is the weight of the terminal edges that $C$ does not cut.
    Adding the weight $\sum_{v, i} g_{v, i} \one{C(v) \ne i}$ of the cut terminal edges to both sides of the chain gives $\cost{C} \ge \sum_{v, i} g_{v, i} - R = B - R$, by~\eqref{eq:graphcost}.

    For the second assertion, let $C$ be a random non-opposite cut of $\Delta_q$ with approximation factor $\alpha$.
    Restricted to $V$, every realization is a non-opposite cut of $H$, so $\cost{C} \ge B - R$ holds for every realization.
    Every edge $uv$ is cut with probability at most $\alpha \, d(u, v)$, and every terminal edge $v e_i$ with probability at most $\alpha \, d(v, e_i) = \alpha (1 - v_i)$, because $C(e_i) = i$.
    Taking expectations in~\eqref{eq:graphcost} gives
    \[
        B - R \le \EE{\cost{C}} \le \alpha \sum_{uv \in E} w_{uv} \, d(u, v) + \alpha \sum_{v, i} g_{v, i} (1 - v_i) = \alpha L,
    \]
    and dividing by $L > 0$ gives $\alpha \ge (B - R)/L$.
\end{proof}

For a small calculation, take the path $a b c$ at $a = (0, 1/2, 1/2)$, $b = (1/3, 1/3, 1/3)$, and $c = (2/3, 1/6, 1/6)$, with weights $w_{ab} = 2$ and $w_{bc} = 1$.
Set $\varphi_1(a \to b) = 2$ and $\varphi_1(b \to c) = 1$, and set every other label flow to zero.
Choose terminal weights $g_{a, 2} = 1/2$, $g_{b, 1} = g_{c, 1} = 1$, all others zero, and allowances $\epsilon_a = 1/2$, $\epsilon_b = \epsilon_c = 0$.
Capacity holds, and $\operatorname{in}_1(a) = - 2$, $\operatorname{in}_1(b) = 2 - 1 = 1$, and $\operatorname{in}_1(c) = 1$.
The negative inflow at $a$ is permitted because label $1$ is forbidden there.
The allowance at $a$ covers $g_{a, 2}$, since $\operatorname{in}_2(a) = 0$.

Label the vertices $C(a) = \ast$ and $C(b) = C(c) = 1$.
The selected inflows are $0, 1, 1$.
The uncut edge $bc$ contributes $1 - 1 = 0$, whereas the cut edge $ab$ contributes $2 - 0 = 2$.
The terminal edges from $b$ and $c$ to $e_1$ are not cut, with total weight $2$, and the terminal edge from $a$ to $e_2$ is cut, with weight $1/2$.
Thus $B = 5/2$, $R = 1/2$, and
\[
    L = 2 \cdot \frac{1}{3} + 1 \cdot \frac{1}{3} + \frac{1}{2} \cdot \frac{1}{2} + 1 \cdot \frac{2}{3} + 1 \cdot \frac{1}{3} = \frac{9}{4},
\]
where the first two terms in $L$ are the path edges and the last three are the terminal edges.
The cut costs $2 + 1/2 = 5/2 \ge B - R = 2$.

\begin{proof}[Proof of \cref{prop:discretelimit}]
    Let the label flows $\varphi_\ell$ and the allowances $\epsilon_v$ form a flow instance on $H = (V, E)$, with $g_{v, i} = 0$ for $i \notin \supp(v)$ as above.
    At $v$, multiply the conservation inequality $\operatorname{in}_i(v) \ge g_{v, i} - \epsilon_v$ for label $i$ by $v_i$ and sum over $i \in [q]$.
    These are allowed labels whenever $v_i > 0$, and $\sum_i v_i = 1$, so
    \[
        \sum_{v, i} v_i \operatorname{in}_i(v) \ge \sum_{v, i} v_i g_{v, i} - R.
    \]
    An edge $uv$ contributes $\sum_i (v_i - u_i) \varphi_i(u \to v)$ to the left side, since it appears as $v_i \varphi_i(u \to v)$ in the sum at $v$ and as $u_i \varphi_i(v \to u) = - u_i \varphi_i(u \to v)$ in the sum at $u$.
    The coefficients $v_i - u_i$ sum to zero, their positive and negative masses both equal $d(u, v)$, and they vanish unless $i \in \supp(u) \cup \supp(v) \subseteq A(u) \cup A(v)$.
    By capacity the values $\varphi_i(u \to v)$ for these labels lie in an interval of length at most $w_{uv}$, so the sum is at most $d(u, v)$ times the length of that interval, and
    \[
        \sum_i (v_i - u_i) \varphi_i(u \to v) \le d(u, v) \, w_{uv}.
    \]
    Summing over edges and adding $\sum_{v, i} g_{v, i} (1 - v_i)$ yields
    \[
        L \ge \sum_{v, i} v_i g_{v, i} - R + \sum_{v, i} (1 - v_i) g_{v, i} = B - R. \qedhere
    \]
\end{proof}

The obstruction concerns label flows on edges, not finite graphs themselves, since the proof applies to the fractional labeling $p_v(i) = v_i$, in which each edge compares its endpoint distributions independently.
On the simplex, capacity is imposed on the surfaces along which the label of a cut changes, which the fractional solution does not have (\cref{sec:overview-simplex}), and \cref{sec:proof} proves the resulting bound.

\section{Verification of the Triangle Flow Instance}\label{app:triangle}

\Cref{sec:triangle} read the three conditions of \cref{def:certificate} off the figure for the flow instance on the triangle and computed its three numbers.
This appendix verifies the conditions at every point, with the conventions of \cref{sec:construction}, and so completes the proof of \cref{prop:triangle}.

\begin{proof}[Proof of \cref{prop:triangle}]
    \emph{Normal continuity.}
    Every face is one cell, so there is nothing to check.

    \emph{Terminal condition and supply.}
    On $\Phi^{[3]}_i$ we have $p^{[3]}_i = 0$, so the $i$th component of $F^{[3]}_\ell$ is $\one{\ell = i} = g^{[3]}_i \one{\ell = i}$.
    By~\eqref{eq:facetnormals} this is the outward flux, so label $i$ leaves with density $1$ and the other labels do not leave.
    At the endpoint $x_i = 2/3$ of $\Omega_{\set{i, j}}$, the $i$th component of $F^{\set{i, j}}_i$ is $(2/3)(2/3 - 1/3) = 2/9 = g^{\set{i, j}}_i$, while $F^{\set{i, j}}_j$ has $x_j = 1/3$ there and vanishes, as does $F^{\set{i, j}}_\ast$.
    On the opposite side $x_i = 0$ of the hexagon, the $i$th component of $F^{[3]}_i$ is $1 - 2/3 = 1/3$, so this flow enters the hexagon where label $i$ is forbidden.
    For the supply into the interval $\Omega_{\set{i, j}}$, let $s$ be the missing coordinate.
    By~\eqref{eq:supply},
    \[
        b^{[3] \to \set{i, j}}_i = - F^{[3]}_{i, s} \big|_{x_s = 0} = - \Bigl( 0 - \frac{2}{3} + 0 \Bigr) = \frac{2}{3},
    \]
    and likewise $b^{[3] \to \set{i, j}}_j = 2/3$ and $b^{[3] \to \set{i, j}}_\ast = 0$.

    \emph{Capacity.}
    On the hexagon the nonempty proper subsets of $[3]$ are the singletons and their complements, and since every flow is tangent the inequality~\eqref{eq:cap} for $I$ and for $[3] \setminus I$ coincide.
    For $I = \set{i}$ the right side is $w^{[3]}_{ij} + w^{[3]}_{is} = 1$.
    The $i$th component of $F^{[3]}_\ell$ is $1 - p^{[3]}_i$ for $\ell = i$, $- p^{[3]}_i$ for the other two terminal labels, and $0$ for $\ast$.
    Since $0 \le p^{[3]}_i \le 2/3$ on the hexagon, any two of these values differ by at most $1$.
    In the normal form~\eqref{eq:normalbound}, this reads
    \[
        \frac{1}{2} \bigl( \abs{\nu_1 - \nu_2} + \abs{\nu_1 - \nu_3} + \abs{\nu_2 - \nu_3} \bigr) = \max_s \nu_s - \min_s \nu_s \ge \abs{\nu(F^{[3]}_\ell - F^{[3]}_m)},
    \]
    since $\nu(F^{[3]}_i - F^{[3]}_j) = \nu_i - \nu_j$ and $\nu(F^{[3]}_i - F^{[3]}_\ast) = \nu_i - \sum_s p^{[3]}_s \nu_s$ is the difference between $\nu_i$ and a weighted average of the $\nu_s$.

    On the interval $\Omega_{\set{i, j}}$ the only subset is $I = \set{i}$, with right side $w^{\set{i, j}}_{ij} = 2/9$.
    Since $x_i + x_j = 1$, the $i$th components of the two terminal flows are $(2/3)(x_i - 1/3) \in [0, 2/9]$ and $- (2/3)(x_j - 1/3) = (2/3)(x_i - 2/3) \in [- 2/9, 0]$, with difference
    \[
        \frac{2}{3} \Bigl( x_i - \frac{1}{3} \Bigr) + \frac{2}{3} \Bigl( x_j - \frac{1}{3} \Bigr) = \frac{2}{3} \Bigl( x_i + x_j - \frac{2}{3} \Bigr) = \frac{2}{9},
    \]
    and the component of $F^{\set{i, j}}_\ast$ is $0$, between them.
    So every pair differs by at most $2/9$.

    \emph{Conservation.}
    In the coordinates $x_1, x_2$ on the hexagon, the components of a terminal flow are $F^{[3]}_{\ell, 1} = \one{\ell = 1} - 2/3 + x_1$ and $F^{[3]}_{\ell, 2} = \one{\ell = 2} - 2/3 + x_2$, so
    \[
        \divergence F^{[3]}_\ell = \frac{\partial F^{[3]}_{\ell, 1}}{\partial x_1} + \frac{\partial F^{[3]}_{\ell, 2}}{\partial x_2} = 1 + 1 = 2
        \text{ and }
        \divergence F^{[3]}_\ast = 0.
    \]
    The top face has no parent, and the allowance $2$ covers this divergence.
    On $\Omega_{\set{i, j}}$, in the coordinate $x_i$, the component of $F^{\set{i, j}}_i$ is $(2/3)(x_i - 1/3)$ and that of $F^{\set{i, j}}_j$ is $(2/3)(x_i - 2/3)$, both with derivative $2/3$, so
    \[
        \divergence F^{\set{i, j}}_i = \divergence F^{\set{i, j}}_j = \frac{2}{3} = b^{[3] \to \set{i, j}}_i = b^{[3] \to \set{i, j}}_j
        \text{ and }
        \divergence F^{\set{i, j}}_\ast = 0 = b^{[3] \to \set{i, j}}_\ast,
    \]
    and conservation holds on the intervals with zero allowance.
    The hexagon supplies exactly the growth of the interval flows, which is the balance that fixed the interval slope.

    \emph{The three numbers.}
    They were computed in \cref{sec:triangle}, where $B = 7/3$, $R = 2/3$, and $L = 3/2$, and the last assertion of the proposition is \cref{thm:general}.
\end{proof}

\end{document}